\documentclass[letterpaper,twocolumn,10pt]{article}
\usepackage{usenix,epsfig,endnotes}
\usepackage{tikz}
\usepackage{amsmath}
\usepackage{amsmath}
\usepackage{amssymb}
\usepackage{bbm} 
\usepackage{xcolor}
\usepackage[ruled,vlined,linesnumbered]{algorithm2e}
\usepackage{enumitem}
\usepackage{listings}
\usepackage{filecontents}
\makeatletter
\let\oldnl\nl
\newcommand{\nonl}{\renewcommand{\nl}{\let\nl\oldnl}}
\makeatother
\usepackage{booktabs}   
\usepackage{multirow}   
\usepackage{array}      
\usepackage{graphicx}   

\newcommand{\sectionref}[1]{\hyperref[#1]{\S\ref*{#1}}}
\newcommand{\figureref}[1]{\hyperref[#1]{Fig. \ref*{#1}}}
\newcommand{\tabref}[1]{\hyperref[#1]{Table \ref*{#1}}}

\usepackage{tikz}
\usepackage{amsmath}
\usepackage{xparse}
\usepackage{caption}
\usepackage{subfig}
\usepackage{setspace}
\usepackage{hyperref}
\usepackage[normalem]{ulem}
\usepackage{lipsum}
\usepackage{listings}
\usepackage{pifont}
\usepackage{algorithmic}

\usepackage[utf8]{inputenc}
\usepackage{listings}
\usepackage{xcolor}

\usepackage{color}
\definecolor{codegreen}{rgb}{0,0.6,0}
\definecolor{codegray}{rgb}{0.5,0.5,0.5}
\definecolor{codepurple}{rgb}{0.58,0,0.82}
\definecolor{backcolour}{rgb}{0.95,0.95,0.92}
\definecolor{textblue}{rgb}{.2,.2,.7}
\definecolor{textred}{rgb}{0.54,0,0}
\definecolor{textgreen}{rgb}{0,0.43,0}
\definecolor{codered}{rgb}{201,72,12}

\usepackage[T1]{fontenc}
\usepackage[scaled=0.85]{beramono} 
\usepackage{listings}
\usepackage{multirow}
\usepackage{setspace}
\usepackage{tikz}

\definecolor{codegreen}{rgb}{0,0.6,0}
\definecolor{codegray}{rgb}{0.5,0.5,0.5}
\definecolor{codepurple}{rgb}{0.58,0,0.82}
\definecolor{backcolour}{rgb}{0.95,0.95,0.92}
\definecolor{textblue}{rgb}{.2,.2,.7}
\definecolor{textred}{rgb}{0.54,0,0}
\definecolor{textgreen}{rgb}{0,0.43,0}
\definecolor{codered}{rgb}{201,72,12}

\usepackage[T1]{fontenc}
\usepackage[scaled=0.85]{beramono} 

\begin{document}

\date{}

\title{\Large \bf An Efficient and Adaptive Watermark Detection System with Tile-based Error Correction}

\author{
\begin{tabular}{c}
Xinrui Zhong, Xinze Feng, Jingwei Zuo, Fanjiang Ye, Yi Mu\textsuperscript{$\diamond$}, \\
Junfeng Guo\textsuperscript{$\dagger$}, Heng Huang\textsuperscript{$\dagger$}, Myungjin Lee\textsuperscript{$\ddagger$}, Yuke Wang \\
Rice University \quad
\textsuperscript{$\diamond$}\,University of Illinois Urbana–Champaign\\
\textsuperscript{$\dagger$}\,University of Maryland \quad
\textsuperscript{$\ddagger$}\,Cisco Research \quad
\end{tabular}
}

\maketitle

\thispagestyle{empty}

\subsection*{Abstract}
Efficient and reliable detection of generated images is critical for the responsible deployment of generative models. Existing approaches primarily focus on improving detection accuracy and robustness under various image transformations and adversarial manipulations, yet they largely overlook the efficiency challenges of watermark detection across large-scale image collections.
To address this gap, we propose \textbf{QRMark}, an efficient and adaptive end-to-end method for detecting embedded image watermarks. The core idea of QRMark is to combine QR Code–inspired error correction with tailored tiling techniques to improve detection efficiency while preserving accuracy and robustness. At the algorithmic level, QRMark employs a Reed–Solomon error correction mechanism to mitigate the accuracy degradation introduced by tiling. At the system level, QRMark implements a resource-aware multi-channel horizontal fusion policy that adaptively assigns more streams to GPU-intensive stages of the detection pipeline. It further employs a tile-based workload interleaving strategy to overlap data-loading overhead with computation and schedules kernels across stages to maximize efficiency. End-to-end evaluations show that QRMark achieves an average $2.43\times$ inference speedup over the sequential baseline.

\section{Introduction} \label{sec:intro}
Recent advances in diffusion-based generative models~\cite{rombach2022high, ho2020denoising, nichol2021improved, song2020denoising} have progressively narrowed the gap between synthetic outputs and real images, making them increasingly difficult to distinguish. In this context, watermarking has become a crucial tool for verifying AI-generated content and supporting copyright attribution. Today, major social platforms handle billions of new visual uploads each day. For example, YouTube reports over 500 hours of video uploaded per minute on its official site~\cite{YouTubeHowItWorksSearch}, and a recent measurement study estimates approximately 117 million TikTok posts per day~\cite{Steel2025TikTokHour}. Historical records also show Facebook receiving more than 350 million new photos per day~\cite{Wang2015FaceSearchAtScale}. The growing scale of online content creation highlights the need for watermark detection pipelines that not only achieve high accuracy, but also maintain robustness against adversarial perturbations and provide high efficiency when processing massive volumes of images.

Among existing watermarking techniques, \textit{in-generation watermarking}~\cite{fernandez2023stable, wen2023tree, wang2025sleepermark}, where the watermark is incorporated directly into the generative process, has received the most attention and has emerged as the most widely adopted solution due to its accuracy and robustness. However, existing efforts largely overlook two critical practical factors of this method: the \textit{latency} of watermark detection and the \textit{throughput} required to reliably handle the massive volume of incoming images at scale. For example, even for relatively small $256 \times 256$ images, the representative Stable Signature model requires about $30$ minutes to process $1$ million images—far from sufficient for social platforms that must process continuous streams of large-scale image and video content.

One potential strategy to reduce computational overhead is tiling, where the detection process operates on a tile that still contains the embedded message. While tiling can substantially lower the computational cost of watermark detection, naive tiling does not necessarily yield proportional performance gains and often results in a noticeable drop in accuracy and robustness. Our experiments show that using the original pretrained Stable Signature decoder for tile-level detection achieves only $87.5\%$ bit accuracy, compared to $99.7\%$ with the full Stable Signature pipeline. Additional experiments consistently confirm that tiling inevitably degrades detection performance, underscoring the intrinsic trade-off between computational efficiency and accuracy. In this context, we require a method that combines low computational overhead with strong error-correction capability to offset the accuracy loss induced by tile-based watermarking. A familiar example is the QR Code, which remains reliably decodable even under partial occlusion or physical damage, thereby ensuring preservation of the stored data. Motivated by this observation, we examined the underlying mechanism and found that this resilience is largely enabled by Reed–Solomon (RS) error correction~\cite{reed1960polynomial,lim1978decoding}. We observe that RS correction holds potential for mitigating the accuracy and robustness degradation caused by tiling, as it can flexibly recover the original message bit string from partially corrupted outputs.
\begin{figure}[t] \small
  \centering
  \includegraphics[width=\linewidth]{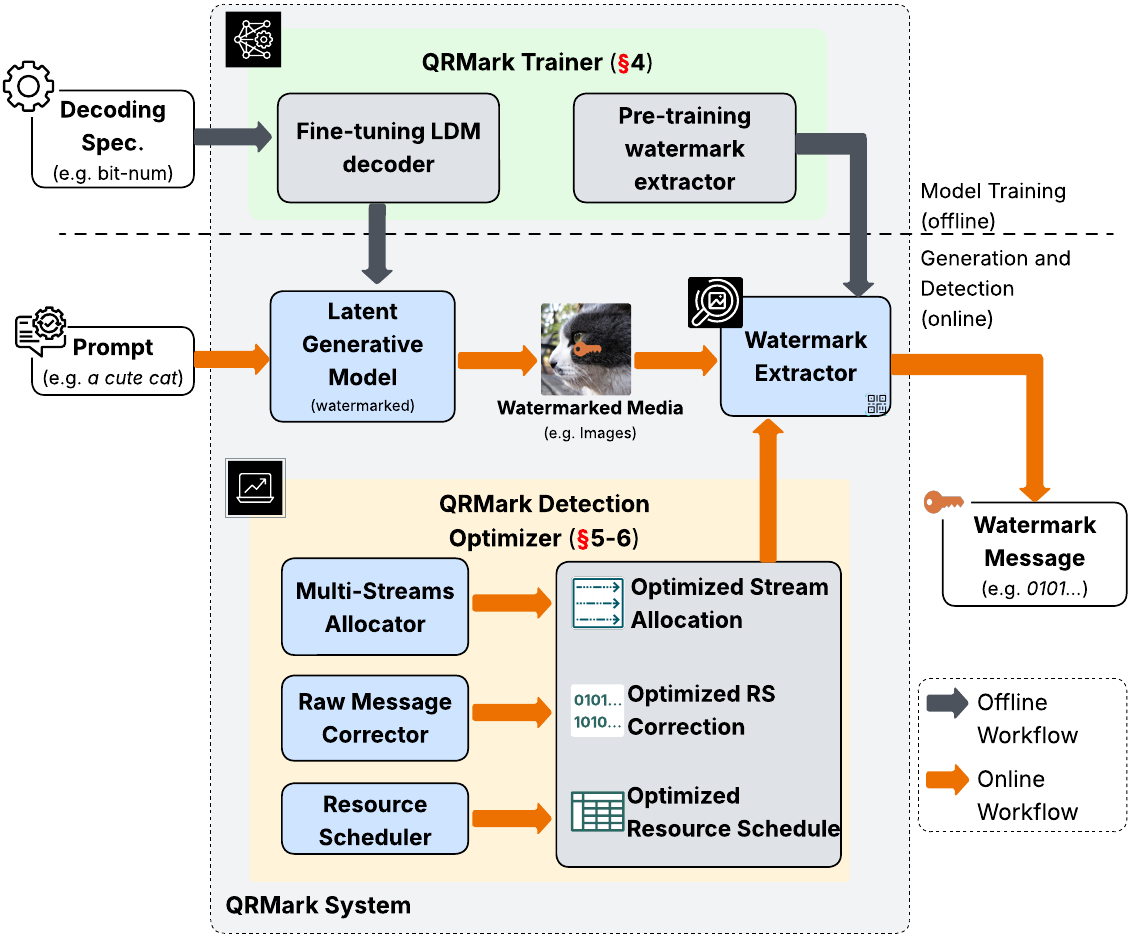}
  \caption{Overview of QRMark. The framework comprises an \textit{offline} stage for training a tile-based watermark extractor and an \textit{online} stage for optimizing the watermark detection pipeline for high-accuracy and low-latency inference.}
  \label{fig:overview}
\end{figure}

Nevertheless, fully harnessing this potential introduces several system-level challenges. Our experiments show that applying a straightforward tiling strategy to Stable Signature yields modest performance gains ($1.17\times$ speedup) but still falls short of expectations. Moreover, RS correction is executed on the CPU, a common design choice since the algorithm is traditionally CPU-bound due to its many interdependent instruction flows (e.g., if–else branches), and thus benefits little from GPU acceleration, which relies on massively parallel execution. In a basic pipeline, this CPU stage can easily become a bottleneck, as it must synchronize with the rest of the detection process running on the GPU. These observations underscore the need to carefully redesign the pipeline to reduce computational overhead and ensure scalability under high-throughput workloads.

To this end, we propose \textbf{QRMark}, an efficient and adaptive watermark detection system with tile-based error correction (\figureref{fig:overview}) for image watermark detection. QRMark combines algorithmic enhancements with system-level optimizations to preserve high detection accuracy and robustness while substantially improving efficiency. At the algorithmic level, QRMark integrates a tile-based decoding strategy with RS correction, achieving a favorable balance between accuracy and efficiency. Specifically, we train a tile-level watermark extractor and employ it during detection to enable parallel processing across images. This restructuring of the original detection pipeline exploits parallelism while maintaining accuracy and robustness, providing a better trade-off between computational complexity and detection accuracy compared to the sequential Stable Signature baseline.

Beyond algorithmic design, QRMark introduces several system-level innovations to maximize GPU utilization and reduce end-to-end latency. First, we implement an adaptive multi-channel horizontal fusion strategy that dynamically adjusts to image resolution and message length, ensuring balanced parallelism without incurring excessive kernel-launch overhead. Second, we design a resource-aware scheduling algorithm for tile-level workloads that distributes computation across GPU streams to mitigate imbalance and improve throughput. Finally, QRMark implements workload interleaving to overlap data loading with GPU execution, alleviating pipeline stall bubbles and enabling more efficient kernel orchestration.

To sum up, we make the following contributions:
\begin{itemize}
\item We introduce a novel method that integrates image tiling with Reed–Solomon (RS) correction (\sectionref{sec:alo}) to effectively balance efficiency–accuracy trade-offs while improving scalability in practical watermark detection.
\item We design an end-to-end tile-based watermark detection framework that builds on this method while also incorporating system-level optimizations, including adaptive multi-channel horizontal fusion(\sectionref{sec:so:opt1}) and resource-aware scheduling (\sectionref{sec:so:opt2}), to effectively overcome the inefficiencies inherent in naive tile-based detection pipelines.
\item Through comprehensive evaluation, we show that QRMark outperforms representative frameworks in throughput and latency, demonstrating its potential for accelerating a wide range of image watermarking systems.
\end{itemize}

\section{Background}

\textbf{Diffusion Watermarking:}  
Diffusion generative models~\cite{dhariwal2021diffusion, ho2020denoising, nichol2021improved, song2020denoising} have renewed interest in watermarking as a means of ensuring content provenance. In this context, watermarking refers to embedding hidden identifiers into AI-generated images, enabling origin verification without relying on visible tags. While traditional methods embed secret pixel patterns~\cite{begum2020digital, wadhera2022comprehensive}, modern approaches integrate watermarking directly into the generative process to improve robustness. Existing image watermarking techniques can be broadly classified into three major categories: \emph{post-generation}, \emph{pre-generation}, and \emph{in-generation}. (1) \emph{Post-generation watermarking}, where the watermark is embedded after the image has been generated~\cite{rahman2013dwt, zhang2019robust}, is straightforward to implement but highly vulnerable in white-box scenarios, as the watermark can be easily removed. (2) \emph{Pre-generation watermarking}, which embeds watermarks into all training images~\cite{yu2021artificial, zhao2023recipe}, incurs prohibitively high computational costs for large-scale diffusion models. (3) \emph{In-generation watermarking}, where the watermark is embedded during the image generation process~\cite{rahman2013dwt, tancik2020stegastamp, zhu2018hidden, luo2020distortion, al2007combined, jing2021hinet, zhang2020udh}, distributes the signal across the entire image, making it difficult to remove without causing noticeable degradation. We adopt \emph{in-generation watermarking} as the focus of our study, since it is widely accepted and provides superior robustness compared to alternative approaches.  

\textbf{System Support for Diffusion Watermarking:}  
Prior system research on diffusion generation has primarily focused on accelerating the long sequential denoising pipeline, introducing optimizations such as kernel fusion, pipeline parallelism, stream batching, and patch parallelism~\cite{tensorrt, li2024distrifusion, kodaira2023streamdiffusion, zhong2023brief}. These methods improve forward sampling efficiency but assume a monolithic, step-driven workload with strong inter-step dependencies, leaving decode-time scheduling demands unaddressed. Interleaving across CPU and GPU stages is crucial for diffusion watermark detection to achieve lower latency, yet this aspect is absent from prior diffusion system designs. System-level efforts in watermarking, on the other hand, have largely targeted post-hoc embedding and detection, typically modeled as simple feed-forward computations with regular GPU kernels and minimal CPU involvement~\cite{hussain2022faststamp, zhang2025robust, uchida2017embedding}. These methods assume image-level operators and homogeneous kernels, making them well-suited for global transforms but incompatible with patch-based workloads. In contrast, diffusion watermark detection relies on patch extraction, fragmenting computation into numerous small kernels and inducing irregular device–host interactions. Therefore, diffusion watermark detection presents new opportunities when treated as an integrated workload rather than two disjoint problems.  
\section{Motivation}
Existing watermarking systems~\cite{fernandez2023stable,wen2023tree,feng2024aqualora,wang2025sleepermark,zhao2023recipe,min2024watermark} largely overlook efficiency considerations, leaving substantial opportunities for further computational optimization. To better understand how watermarking systems can be made more efficient, we focus on a representative model, Stable Signature~\cite{fernandez2023stable}, and systematically explore both algorithmic and system-level optimizations carefully built on top of it.

\begin{figure}[t] \small
  \centering
  \includegraphics[width=0.9\linewidth]{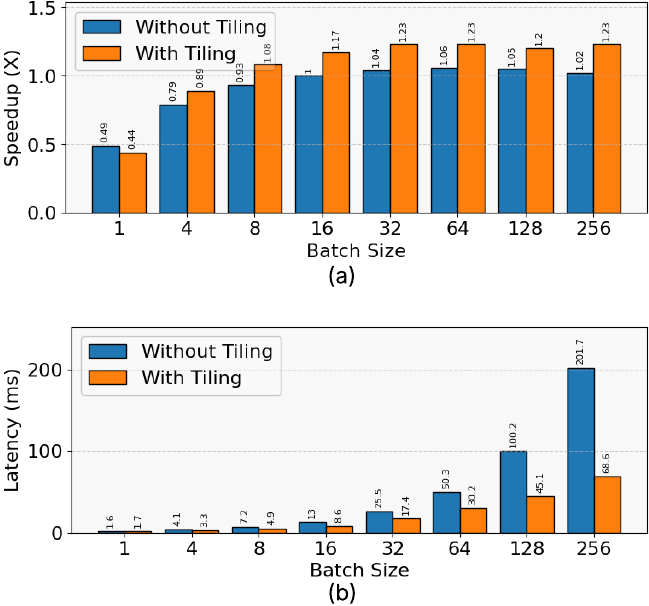}
  \caption{End-to-end performance comparison between Stable Signature and its naive tile-based implementation on (a) speedup and (b) latency.}
  \label{fig:tile_exp}
\end{figure}

\textbf{Divisible image watermark detection with tiling:}  
By analyzing the Stable Signature watermark detection pipeline, we identify an opportunity to exploit tiling. Since an image is inherently divisible, it can be partitioned into tiles for localized processing. Empirical evidence from Stable Signature shows that under a cropping attack where only $10\%$ of the original area is preserved, the model still achieves over $95\%$ detection accuracy~\cite{fernandez2023stable}. This observation suggests that, even without explicit tile-level training, the message extractor is capable of recovering the embedded message from partial image regions. Leveraging this property, we introduce tiling into the detection process. Tiling is a well-established technique in system optimization~\cite{xu2009tiling, rolih2024divide, jangda2020model, ozge2019power}, widely used to reduce both computational and memory overhead. In the context of image watermark detection, our experiment (\figureref{fig:tile_exp}(a)) demonstrates that integrating tiling into Stable Signature yields notable scalability benefits: with tiling, the system achieves up to a $1.23\times$ speedup over the non-tiled baseline, whereas the non-tiled baseline achieves at most $1.06\times$ across varying batch sizes. This result highlights the superior scalability brought by tiling. However, tiling inevitably introduces an efficiency–accuracy trade-off, as partitioning the image may compromise detection fidelity. To address this challenge, we draw inspiration from the design of QR Codes, where Reed–Solomon error correction ensures reliable decoding under data loss and corruption. This motivates our design of QRMark, which integrates Reed–Solomon correction to offset the accuracy degradation caused by tiling and achieve a more favorable balance between efficiency and robustness.

\textbf{Performance dilemma of watermark image tiling:}  
Although replacing full-image processing with tile-level extraction is expected to reduce per-image overhead and improve detection efficiency, our experiments on 1,000 images from the MS-COCO~\cite{lin2015microsoftcococommonobjects} dataset confirm that the naive tile-based design brings only limited improvements: with a tile size of 64 and a batch size of 16, the pipeline achieves a $1.17\times$ speedup over the non-tiled baseline. However, this gain falls short of expectations, prompting the question of what factors constrain the potential of the tile-based approach. To answer this, we conducted profiling experiments with a default tile size of 64, a message length of 48, and a batch size of 16. The results reveal two primary bottlenecks. First, GPU utilization remains low: Stable Signature leverages only $31.6\%$ of available GPU capacity. Second, the Reed–Solomon correction, executed on the CPU, introduces additional overhead and occasionally stalls the entire pipeline. Further experiments also show that the transform phase preceding detection in the original Stable Signature implementation launches a large number of fine-grained kernels. These fragmented kernels originate from preprocessing operations such as cropping, resizing, and normalization. The fragmented design incurs substantial kernel-launch overhead and prevents the GPU from achieving high occupancy due to frequent scheduler invocations.  

\begin{figure}[t]
  \centering
  \includegraphics[width=\linewidth]{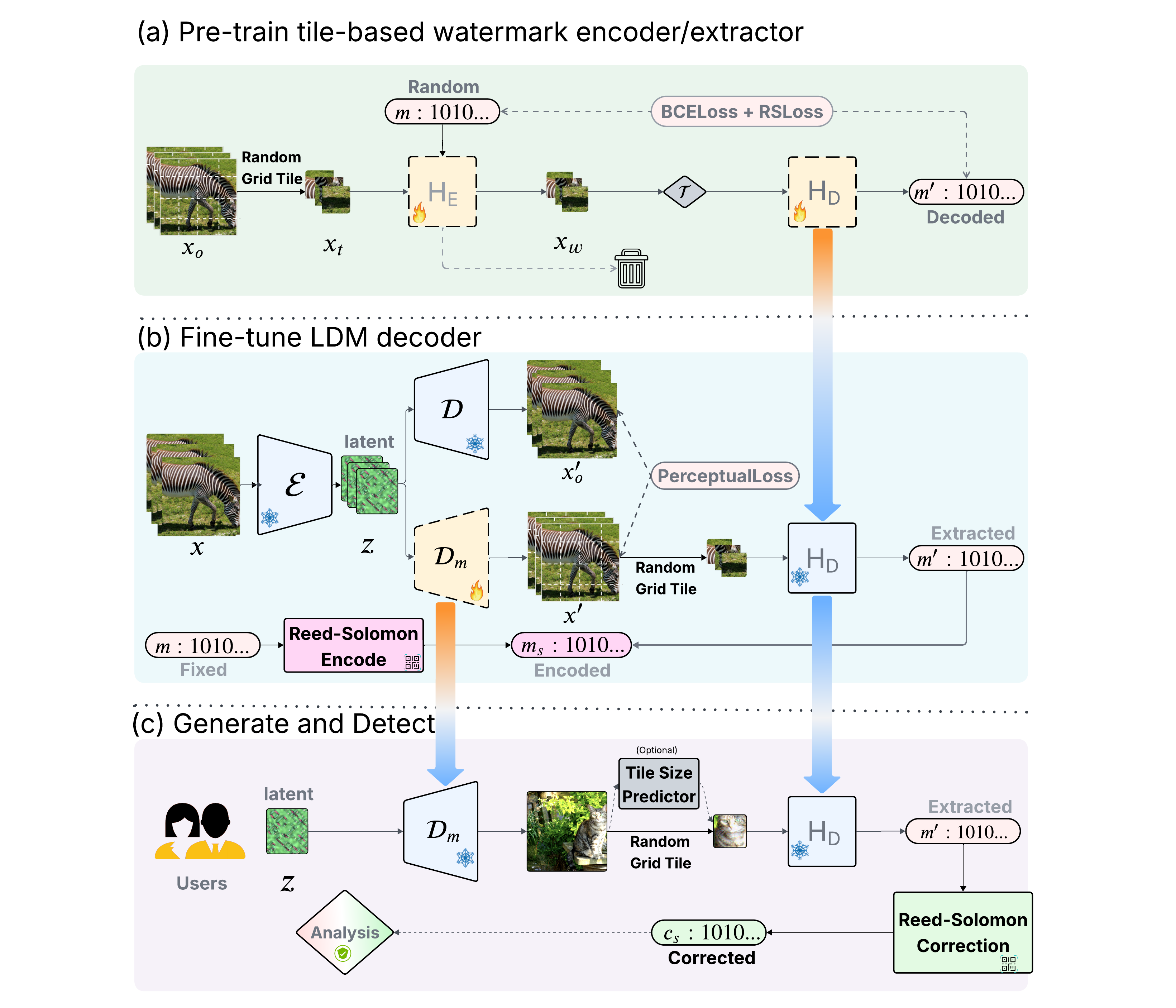}
  \caption{
  Overview of QRMark Algorithm. 
  (a) We pretrain a watermarking encoder $H_E$ and an extractor $H_D$ with customized tiling settings to embed and extract a fixed bit sequence $m$, the encoder $H_E$ is then discard; 
  (b) We fine-tune latent diffusion model's decoder $D_m$ with a Reed–Solomon encoded bit sequence $m_s$;
  (c) We apply the extractor $H_D$ to extract the hidden watermark message from the tiles of generated images. A more detailed breakdown on the detection pipeline is shown in \figureref{fig:opt_overview}.
  }
  \label{fig:steps_of_the_algorithm}
\end{figure}
\textbf{Dynamicity of the tile-based watermark pipeline:}  
One way to better utilize GPU resources is to increase the batch size in the tile-based detection pipeline. However, while larger batches improve throughput, they almost inevitably increase latency. \figureref{fig:tile_exp} shows that simply enlarging the batch size in the naive tile-based pipeline quickly drives throughput to saturation, after which it plateaus or even slightly declines, while latency rises sharply. Moreover, most computation is concentrated in the \emph{decoding} stage, i.e., extracting the watermark from each tile, which overlaps poorly with other operations. Thus, decoding emerges as the primary bottleneck, limiting the benefits of larger batch sizes. To further analyze pipeline behavior, we evaluated different batch sizes under a fixed multi-channel horizontal fusion strategy across stages. For a batch size of 256, setting the stream counts for preprocessing, decoding, and RS correction to $(1,1,16)$ yields a $1.43\times$ latency reduction compared to the single-stream baseline. In contrast, applying the same configuration to a batch size of 16 results in a slowdown, achieving only $0.86\times$ due to kernel-launch overhead. This discrepancy demonstrates that an optimal configuration for one setting does not necessarily transfer to another. These findings highlight the need for a strategy that can adapt to varying tile and batch sizes. Designing such an approach presents two key challenges. First, tile size and batch size alone are insufficient to accurately predict the resource demands of individual stages. Second, simply tuning the number of streams cannot balance stage execution times, often leading to pipeline bubbles or additional launch overhead. An adaptive, resource-aware method is therefore required to balance these competing factors and more effectively exploit GPU.
\section{Tiled Watermark Detection Optimization} \label{sec:alo}
\label{sec:sys:challenges}
While the proposed approach in Stable Signature \cite{fernandez2023stable} achieves high accuracy in watermark recovery, the deployment of the detection pipeline introduces system-level constraints. First, the watermark detection process is compute-intensive: the execution of the message decoder $H_D$ is dominated by kernels that operate on dense pixel data from each watermarked tile. Second, the computational cost is intrinsically tied to the spatial resolution of the inputs. In text-to-image diffusion pipelines, it is standard practice to partition high-resolution images into fixed-size tiles so that watermark encoders and decoders can operate locally on spatial regions\cite{wang2025timestep,ding2024patched, zhang2023sine, madar2025tiled,sun2025patchedserve}. We therefore adapt a similar technique to the watermark detection pipeline. Each time, we only need to use a single tile of the watermarked image instead of the whole image. Under this tiling scheme, the tile size directly determines the workload of these kernels, and increasing the tile dimensions to improve robustness leads to a proportional increase in FLOPs and decoding latency.

Motivated by these constraints, we next describe how QRMark is designed to support efficient and scalable watermark extraction. In this section, we first introduce the pretraining of a decoder to recover message bits from fixed-size image tiles and the fine-tuning of an existing latent diffusion model decoder to embed messages into these tiles. We then present the algorithmic workflow of our message extraction pipeline and conclude by explaining how Reed–Solomon correction is applied to further improve accuracy.

\vspace{5pt}
\subsection{Pre-training Tiled Watermark Extractor}
\label{sec:alo:pre}
The pre-training process is consistent with the approach adopted in Stable Signature~\cite{fernandez2023stable}, with two key modifications: (1) tile-specific adaptations to enable training of a tile-based decoder, and (2) adjustments to the loss function to accommodate Reed–Solomon correction.

In \figureref{fig:steps_of_the_algorithm}(a), the pre-training pipeline consists of a convolutional encoder and a convolutional extractor. The encoder learns to embed imperceptible perturbations into the cover image, while the extractor is simultaneously trained to recover the embedded message from the watermarked image. Both encoder and decoder are optimized by the HiDDeN~\cite{zhu2018hidden}.
At the start of the pre-training, we partition each training image into a regular grid of axis–aligned rectangles of size $l\times l$\, px and randomly sample one grid cell $x_t \in R^{l \times l \times 3}$. Following the approach used in ReDMark~\cite{redmark}, the encoder $H_E$ takes the tile as input along with a $N$-bit binary message $m \in \{0,1\}^N$ and then produces a residual image tile $\delta \in \mathbb{R}^{l \times l \times 3}$, that is 
further multiplied by a scalar hyperparameter $\alpha$ to derive the watermarked image tile $x_w = x_0 + \alpha \delta$.
At every optimization step, we first sample a transform $T$ from a set $\mathcal{T}$ that includes image transformations such as \texttt{jpeg}, and then apply $T$ to the watermarked image tile $x_w$. Eventually, the decoder \(H_D\)  then predicts a \emph{soft} $N$-bit message \(m' = H_D(T(x_w))\) based on the transformed watermarked image $T(x_w)$ as input.
We set the standard binary cross-entropy loss as the message loss,
\[
\mathcal{L}_{m}
  = -\sum_{i=1}^{N}\!
      \bigl[
        m_i\,\log\sigma(m_i') + (1-m_i)\,\log\!\bigl(1-\sigma(m_i')\bigr)
      \bigr],
\]
where $\sigma(\cdot)$ denotes the sigmoid.
To incorporate the Reed-Solomon (RS) correction mechanism introduced later in \sectionref{sec:alo:rs}, we introduce an RS-aware loss that penalizes only bit errors exceeding the correction capacity.  Let $t$ denote the maximum number of symbol errors that an $(n,k)$ RS code can correct, and define the error indicator $e_i = \mathbbm{1}\!\left[\operatorname{sign}(m_i') \neq m_i\right]$.
To make sure errors within the RS capacity ($E \leq t$) incur no additional cost while uncorrectable errors are quadratically penalized, we formulated our RS-aware loss $\mathcal{L}_{\text{RS}}$ as $[\max(0,\;E - t)]^2$ where $E = \sum_{i=1}^{k} e_i$. 
The overall training objective becomes
\[
\min \mathcal{L}
  = \mathcal{L}_{m}
    + \lambda\,\mathcal{L}_{\text{RS}},
\]
where $\lambda$ balances the standard bit error loss and the RS-aware term and is set to $\lambda=1$ in our experiments according to the settings in Stable Signature. 

\subsection{Fine-tuning the Tile-based LDM Decoder}\label{sec:alo:fine}
The fine-tuning strategy closely mirrors that of Stable Signature, while it is adapted to our \emph{tile-based} decoder.  Concretely, we fine-tune the latent-diffusion decoder $\mathcal{D}_{m}$ such that the watermarked image it reconstructs encodes an $N$-bit watermark that can be reliably recovered by the tile extractor $H_D$ from the pre-training phase.
As shown in \figureref{fig:steps_of_the_algorithm}(b), instead of explicitly embedding the fixed message $m$ in the image like what we did in \sectionref{sec:alo:pre}, we first apply the RS encoding scheme (details in \sectionref{sec:alo:rs}) to generate a binary signature $m_s$, with a bit length consistent with the pre-training stage, the $m_s$ consist of original message $m$ and correction symbols $m_c$ used for the RS correction purpose later in the watermark detection phase.
For a training image $\mathbf{x}\in\mathbb{R}^{H\times W\times 3}$, we first obtain its latent $\mathbf{z}= \mathcal{E}(\mathbf{x})\in\mathbb{R}^{h\times w\times c}, h=\tfrac{H}{f},  w=\tfrac{W}{f}$ with the LDM encoder \(\mathcal{E}\) 
where $f$ is a power-of-two down-sampling factor, $c$ is the dimension of channels for the latent code. Then, the decoder $\mathcal{D}_{m}$ transforms $\mathbf{z}$ into a watermarked image
\(
  \mathbf{x}'=\mathcal{D}_{m}(\mathbf{z}),
\)
which is subsequently partitioned into non-overlapping $l\times l$ tiles. We randomly select one of the partitioned tiles and pass it to the extractor $H_D$ to predict the message $\mathbf{m}' = H_D(\mathbf{x}')$.
%
{To constrain distortion, we also reconstruct an unmarked image $\mathbf{x}_{o}'=\mathcal{D}(\mathbf{z})$ with the frozen  decoder $\mathcal{D}$ from the original latent diffusion model}, 
and calculate the Watson-VGG perceptual loss~\cite{czolbe2021lossfunctiongenerativeneural}
  $\mathcal{L}_{\text{i}}=\operatorname{WatsonVGG}(\mathbf{x}',\mathbf{x}_{o}')$.
Thus, the total objective is 
\[\mathcal{L}=\ \mathcal{L}_{\text{m}}\;+\;\lambda_{\text{i}}\;\mathcal{L}_{\text{i}},\]
\label{eq:finetune_loss}
with $\lambda_{\text{i}}=2.0$ unless stated otherwise. 
We fine-tune $\mathcal{D}_{m}$ for $100$ AdamW~\cite{loshchilov2019decoupledweightdecayregularization} iterations
(batch size~$4$) with 20 warm-up iterations to $10^{-4}$ followed by decay to $10^{-6}$. As the parameters here are the same as the best setting in Stable Signature. 

\subsection{Detection with Reed–Solomon Correction}
\label{sec:alo:rs}
During the watermark detection illustrated in \figureref{fig:steps_of_the_algorithm}(c), each candidate image is cropped and resized to $256 \times 256$. The image is then partitioned into an $l \times l$ grid, where $l$ is either specified directly or predicted by the tile-size predictor described in Appendix~\ref{subsec:ml_tile_predictor}. From this grid, a single tile suffices for detection: the selected tile is passed through the decoder $H_D$, which outputs a raw bitstream $m'$.
Because the tile-based strategy processes only a portion of the image, the decoded bitstream may contain more errors than in the full-image Stable Signature model. To address this, Reed–Solomon (RS) error correction is applied to $m'$ to obtain the final watermark message $c_s$, which is compared against the ground-truth $g$.

RS correction operates at the symbol level, with the maximum number of correctable errors given by $t = \left\lfloor \tfrac{m_c}{2} \right\rfloor$, where $m_c$ denotes the number of redundant correction symbols. In practice, we set the symbol size to 8 bits and use $m_c = 2$, thereby enabling correction of up to one erroneous symbol. A systematic Berlekamp-Welch (B-W) decoder is then employed to reconstruct the original message from the noisy bitstream. Empirically, we observe that our training and fine-tuning pipeline yields high per-tile extraction accuracy ($>$90\%) for most tile size settings. As a result, correcting a single erroneous symbol is sufficient to recover the small residual errors introduced during decoding and prevents further accuracy degradation in practice. Finally, the RS-corrected output is compared with the predetermined key for robust statistical verification. We provide full details of the RS error-correction procedure in Appendix~\ref{sec:rs}.

\section{Watermark Pipeline Optimization}\label{sec:so:opt1}

In this section, we first present the image watermark detection pipeline, followed by the adaptive multi-channel horizontal fusion algorithm, which jointly optimizes stream allocation and batch partitioning, together with co-optimization of RS correction.

\label{subsec:image_operations}
\begin{table}[t]
  \centering
  \caption{\raggedright Image operations used by QRMark on Stable Signature across different stages.}
  \vspace{2px}
  \label{tab:qrmark_operation}
  \setlength{\tabcolsep}{3pt}
  \footnotesize
  \resizebox{\columnwidth}{!}{
    \begin{tabular}{cll}
      \toprule
      \textbf{Stage} & \textbf{Operation} & \textbf{Description} \\ \midrule
      \multirow{4}{*}{Preprocess}
        & \texttt{resizeto($x$)}       & Resize image to target resolution $x$ \\
        & \texttt{centercrop($x$)}     & Crop central patch to target size $x$ \\
        & \texttt{toTensor}            & Convert image to PyTorch tensor \\
        & \texttt{normalize image}     & Normalize tensor to VQGAN range \\ \midrule
      \multirow{3}{*}{Tile}
        & \texttt{random}              & Randomly sample a complete $n \times n$ tile $x_t$  from $x_0$ \\
        & \texttt{random grid}         & Randomly Sample $x_t$ from a size-aligned grid cell \\
        & \texttt{fixed}               & crop $x_t$ from the top-left corner of $x_0$ \\ 
        \midrule
      \multirow{7}{*}{Evaluation}
        & \texttt{crop($x$)}           & Center crop removes image borders \\
        & \texttt{resize($x$)}         & Resize image to $x$ of the original size ($0 < x < 1$) \\
        & \texttt{brightness($x$)}     & Adjust brightness to $x$ times the original intensity \\
        & \texttt{contrast($x$)}       & Adjust contrast to $x$ times the original dynamic range \\
        & \texttt{saturation($x$)}     & Adjust color saturation to $x$ times the original level \\
        & \texttt{sharpness($x$)}      & Adjust sharpness to $x$ times the original edge strength \\
        & \texttt{overlay text}        & Overlay fixed text to simulate occlusion artifacts \\
        \bottomrule
    \end{tabular}
  }
\end{table}

\subsection{Image Watermark Detection Pipeline} \label{subsec:so:pipe}

Real-world watermark detection using QRMark often involves the following stages: \emph{preprocessing}, \emph{tiling}, and \emph{decoding}. We detail each of these stages as follows.

\begin{figure*}[t]
  \centering
  \includegraphics[width=\linewidth]{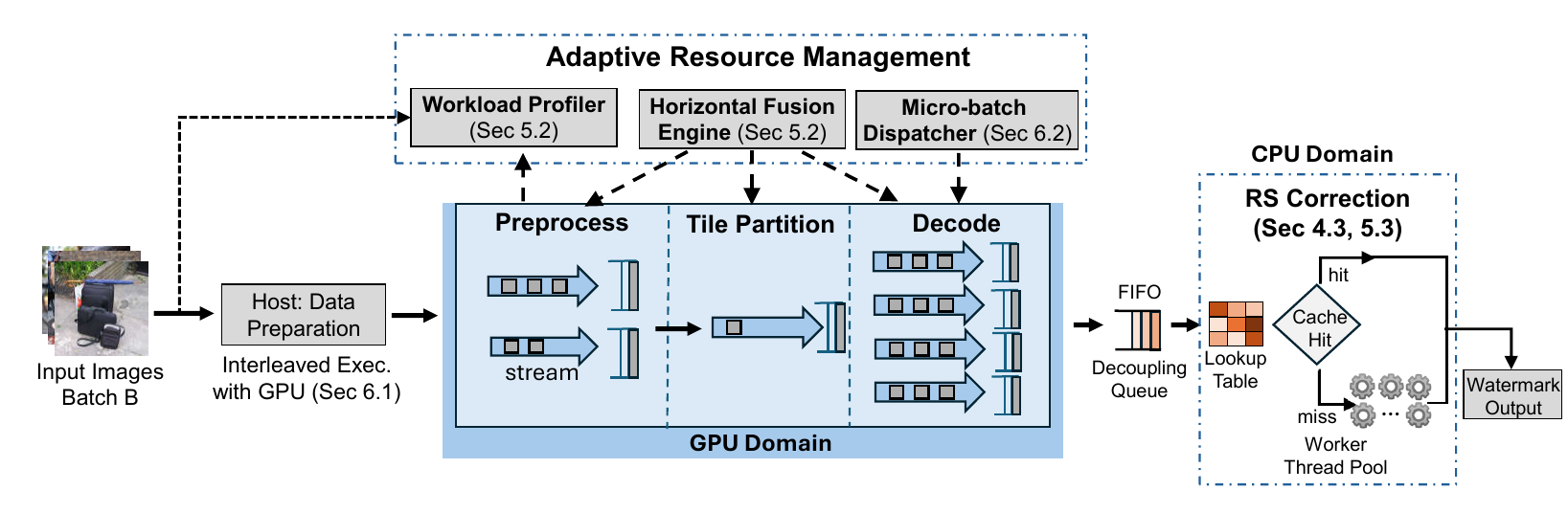}
  \vspace{-25pt}
  \caption{Illustration of optimizations applied in QRMark with multi-channel horizontal fusion and scheduling. Input images are preprocessed and partitioned into tiles on the GPU, followed by multi-channel decoding coordinated by adaptive resource management. Decoded messages are passed through a decoupled FIFO to the CPU-side Reed–Solomon correction for final watermark recovery.}
  \vspace{-10pt}
  \label{fig:opt_overview}
\end{figure*}

\begin{figure}[t]
  \centering
  \includegraphics[width=\linewidth]{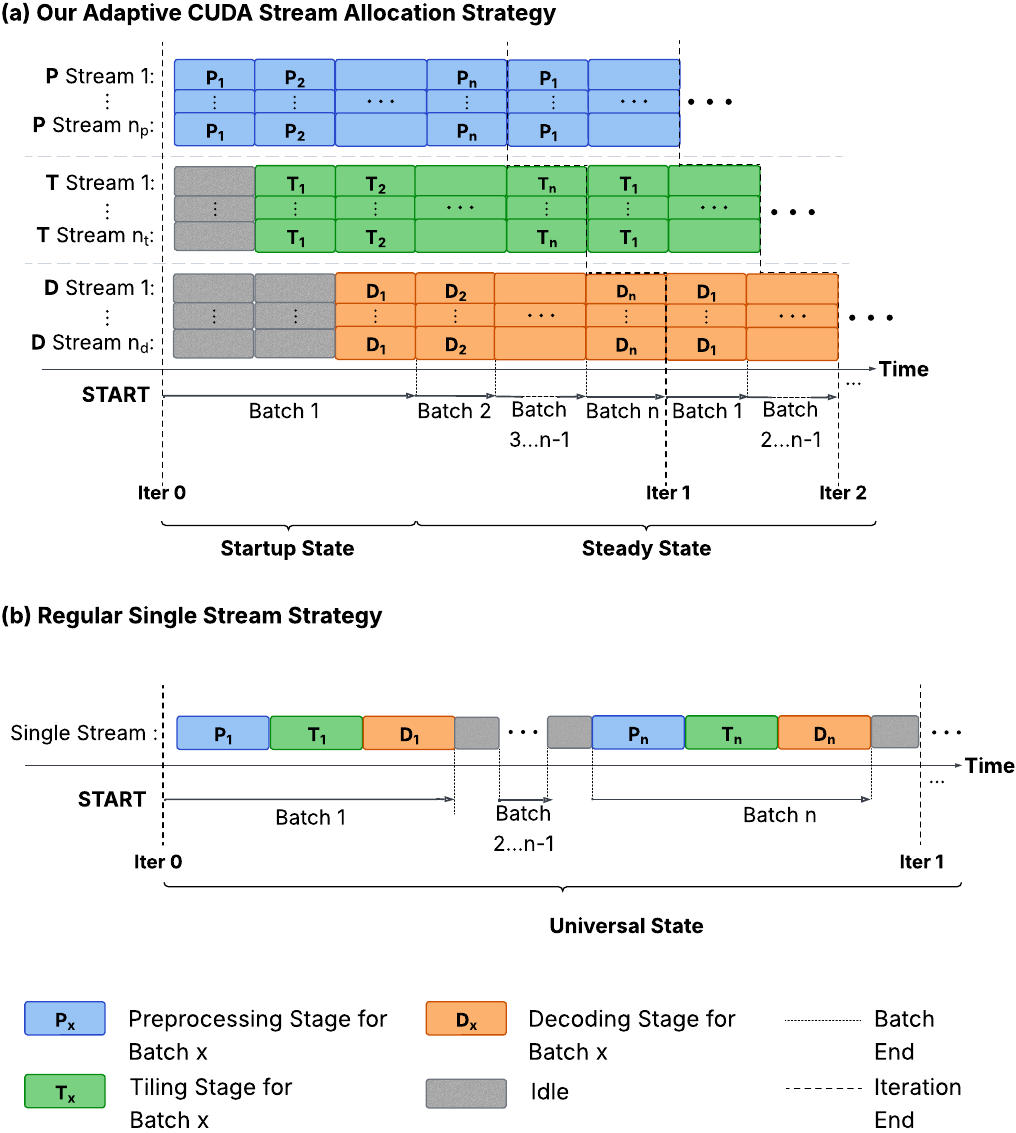}
  \caption{Comparison of multi-channel horizontal fusion strategies in the QRMark image watermark detection pipeline.
(a) \emph{multi-channel horizontal fusion} dynamically distributes streams across preprocessing, tiling, and decoding stages based on measured GPU utilization, thereby enabling greater stage overlap and reducing idle time.
(b) \emph{Single Stream Strategy} executes all stages strictly sequentially, often resulting in noticeable pipeline stalls during CPU-side RS correction.}
  \label{fig:workflow}
\end{figure}

We show the optimizations applied to the detection pipeline in \figureref{fig:opt_overview}. The \emph{preprocessing} stage consists of two substages: data loading and image transformation. As summarized in \tabref{tab:qrmark_operation}, QRMark applies a sequence of image processing operations to Stable Signature. The data loading substage transforms the raw image into a tensor and transfers it to the GPU. The transformation substage then performs \emph{CenterCrop (256)}, \emph{Resize (256)}, and \emph{Normalization}. In the \emph{tiling} stage, possible tiling strategy includes: \texttt{random}, \texttt{random\_grid}, and \texttt{fixed}. We use \texttt{random\_grid} as our default tiling strategy because it demonstrates higher robustness. Each watermarked image is divided into a regular grid composed of axis-aligned rectangular regions, each of size $n\times n$ pixels. A single grid cell is then randomly selected as the input for the \emph{decoding} stage.
In the \emph{decoding} stage, the selected image tensor is processed by a pretrained extractor to recover the raw message bits. The raw message is then passed through our RS error correction module to obtain the final watermark. To mitigate the bottleneck introduced by RS correction, we employ an input queue together with a CPU-based thread pool. In addition, as described in \sectionref{sec:so:rs}, we introduce a caching mechanism to further reduce computational overhead. The CPU thread pool scales nearly linearly with the thread count $t$; in practice, we set $t=32$, which is sufficient for most workloads.

\begin{algorithm}[t]\small
\renewcommand{\baselinestretch}{0.9}\selectfont
\SetAlgoNlRelativeSize{-1}
\SetAlCapSkip{0.5ex}
\caption{Adaptive Multi-channel Horizontal Fusion.}
\label{alg:asa_simple}
\KwIn{Warm-up iters $w$; per-stage time $t[k]$; per-sample memory $u[k]$; global batch $B$; stream budget $P$; memory cap $M_{\mathrm{cap}}$; threshold $\epsilon$; stall cap $\tau$}
\KwOut{Streams per stage $s[1..K]$, micro-batches per stage $m[1..K]$}

\nonl\textbf{\textcolor{blue}{/* Step 1: Warm-up profiling */}}\\
Run $w$ iterations to estimate $t[k]$ and $u[k]$\;
Initialize $s[k]\leftarrow 1$ for all $k$\;
Pick the largest uniform $m$ s.t.\ $\sum_k s[k]\cdot m\cdot u[k]\le M_{\mathrm{cap}}$, set $m[k]\leftarrow m$\;
$J^\star \leftarrow \max_k \textsc{Time}(k,s[k],m[k])$;\quad $\text{stall}\leftarrow 0$\;
\nonl\textbf{\textcolor{blue}{// $J^\star$ = current bottleneck latency across stages}}\\

\nonl\textbf{\textcolor{blue}{/* Step 2: Adaptive search */}}\\
\While{$\text{stall}<\tau$}{
  $\text{gain}\leftarrow 0$;\quad $\text{best}\leftarrow (s,m)$\;

  \For{$k=1$ \KwTo $K$}{
    $s' \leftarrow s$\; $s'[k]\leftarrow s[k]+1$\;
    \If{$\textsc{MemOK}(s',m,M_{\mathrm{cap}})$}{
      $J' \leftarrow \max_j \textsc{Time}(j,s'[j],m[j])$\;
      $\Delta \leftarrow J^\star - J'$\;
      \If{$\Delta>\text{gain}$}{ $\text{gain}\leftarrow \Delta$\; $\text{best}\leftarrow (s',m)$\; }
    }
  }

  \eIf{$\text{gain}>\epsilon$}{
    $(s,m)\leftarrow \text{best}$\;
    $J^\star \leftarrow \max_k \textsc{Time}(k,s[k],m[k])$\;
    $\text{stall}\leftarrow 0$\;
  }{
    $\text{stall}\leftarrow \text{stall}+1$\;
  }
}

\nonl\textbf{\textcolor{blue}{/* Step 3: micro-batch leveling */}}\\
$u_s \leftarrow \sum_k s[k]$;\quad $m_{\text{unit}} \leftarrow \max(1,\lfloor B/u_s \rfloor)$\;
\For{$k=1$ \KwTo $K$}{
  \If{$\textsc{Time}(k,s[k],m[k]) \ll J^\star$ \textbf{ and } memory allows}{
    $m[k]\leftarrow \min(m_{\text{unit}},2m[k])$\;
  }
}
\Return{$(s,m)$}\;
\end{algorithm}

\subsection{Adaptive Multi-channel horizontal fusion} \label{sec:so:sso:plp}
Inspired by the minibatch design in PipeDream~\cite{narayanan2019pipedream}, we propose a multi-channel horizontal fusion strategy in \figureref{fig:workflow} to mitigate preprocessing and RS correction bottlenecks in the detection pipeline. Our primary design goal is to equalize the per-minibatch execution time across stages, thereby maximizing effective overlap and minimizing pipeline bubbles. Concretely, we partition GPU streams into three groups, each mapped to a pipeline stage. If stage $i$ receives $s_i$ streams, the global batch $B_{\text{cap}}$ is evenly divided into $s_i$ minibatches of size $B_{\text{cap}}/s_i$, and each minibatch is executed on an assigned CUDA stream. The number of streams per stage is chosen adaptively based on warm-up profiling that estimates each stage’s baseline time $t[i]$ and per-sample memory footprint $u[i]$. We allocate more streams to slower stages (large $t[i]$) to contract their step time, subject to a device-memory feasibility constraint $\sum_i s_i \cdot m_i \cdot u[i] \le M_{\mathrm{cap}}$, where $m_i$ is the stage’s micro-batch size. We also enforce a global stream cap $\sum_i s_i \le P$, trading off finer-grained balancing against launch costs.

As detailed in Algorithm~\ref{alg:asa_simple}, the horizontal fusion strategy begins with a warm-up phase to estimate each stage’s execution time and per-sample memory footprint (Line 1). It initializes the configuration by assigning one stream per stage and selecting the largest uniform micro-batch that fits the memory budget (Lines 2--4). It then computes the initial bottleneck latency $J^\star$ and resets the stall counter (Line 5). The algorithm enters the adaptive search loop (Line 8). Within each iteration, it evaluates stream augmentation: for every stage $k$, it tentatively increases $s[k]$ by one, and records the candidate configuration that yields the largest reduction in bottleneck latency (Lines 10--15). After scanning all stages, the algorithm decides whether to accept this best modification: if the improvement exceeds the threshold $\epsilon$, it updates $(s,m)$, recomputes $J^\star$, and resets the stall counter; otherwise, the stall counter is incremented (Lines 16--20). The search terminates once no further gains are observed or the stall counter reaches its cap. After the search halts, the algorithm performs micro-batch leveling: it derives a unit micro-batch $m_{\text{unit}}$ from the total stream count, and for stages substantially faster than the bottleneck, it increases their micro-batch size up to this bound (Lines 22--25). Finally, the algorithm returns the per-stage stream allocation and micro-batch sizes (Line 26).

\subsection{Co-Optimizations for RS Correction}\label{sec:so:rs}

While adaptive multi-channel horizontal fusion alleviates GPU-side contention, the detection pipeline can still suffer from stalls introduced by the Reed–Solomon (RS) correction stage. To address this, we apply optimizations at both the system and algorithmic levels. At the system level, we decouple RS correction from the GPU pipeline by introducing an input queue and a CPU thread pool. In \sectionref{sec:alo:rs}, after the raw watermark messages $m'$ are decoded, $m'$ are dispatched to idle CPU threads for correction, and the final corrected outputs $c_s$ are collected asynchronously. This design prevents device-to-host transfers and CPU computation from stalling GPU progress. At the algorithm level, we observe that the embedded message sets are limited and detection accuracy is usually above 95\%, leading to frequent recurrence of raw messages $m'$. Inspired by that, we propose to maintain a codebook $cb$ that maps each $m'$ to its corrected output $c_s$, together with a counter $c$ that tracks the number of images processed since its last access. When $m'$ reappears, the cached result is directly reused, bypassing the costly correction step. The combination of system-level decoupling and algorithmic caching substantially reduces RS overhead while preserving correction fidelity.
\section{Resource-aware Scheduling Optimization}\label{sec:so:opt2}

In this section, we will demonstrate the inter-batch interleaving to overlap CPU-side data preparation and a resource-aware micro-batch schedule algorithm to balance workloads among streams and improve overall throughput.

\subsection{Inter-Batch Workload Interleaving}

The preprocessing stage in \textsc{QRMark} comprises two substages: CPU-side data preparation and GPU-side preprocessing. Inspired by the interleaving strategy in RAP~\cite{wang2024rap}, we adopt a similar approach that overlaps CPU preparation of the next batch with concurrent GPU kernel execution of the current batch, thereby achieving a more balanced overall schedule and reducing end-to-end latency.

\begin{figure}[t]
  \centering
  \includegraphics[width=\linewidth]{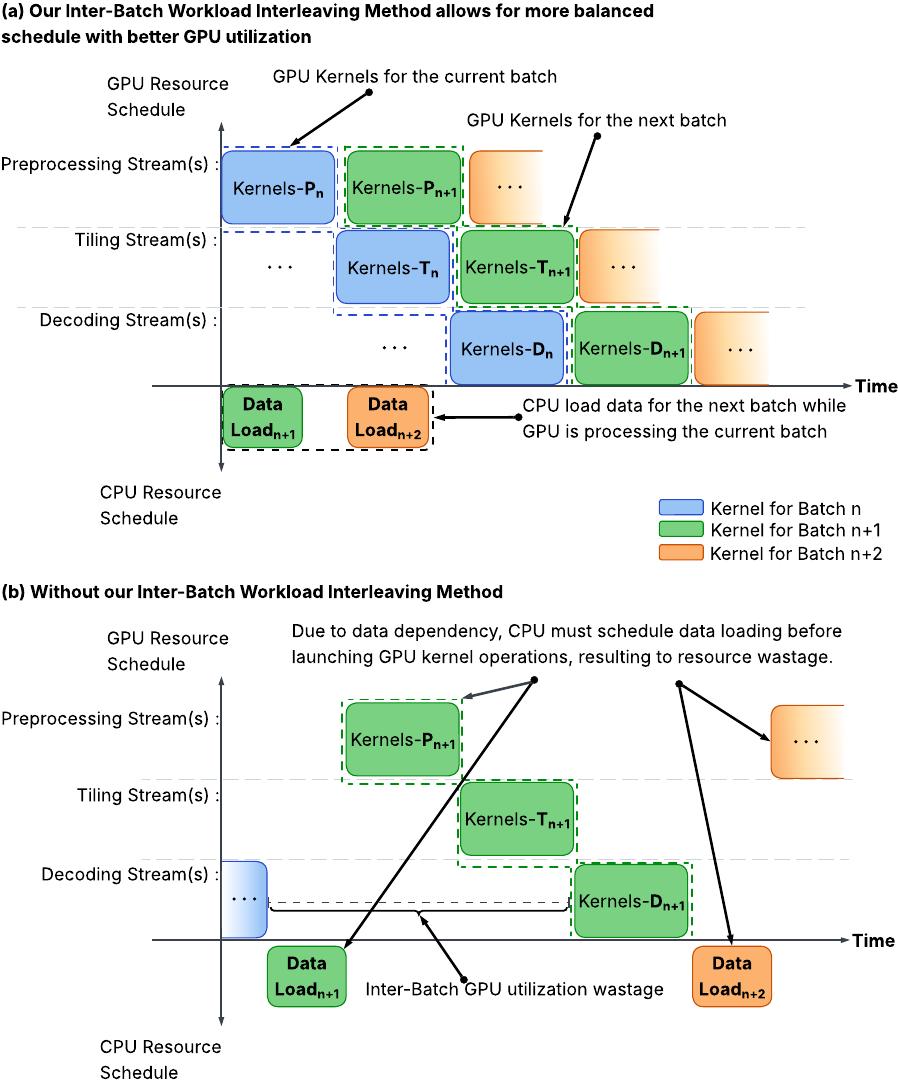}
  \caption{Comparison of resource schedule with and without our Inter-batch Workload Interleaving Strategy.}
  \label{fig:interleaving}
\end{figure}

As illustrated in \figureref{fig:interleaving}, without inter-batch interleaving, preprocessing must fully complete before subsequent stages (tiling and decoding) can begin, due to strict data dependencies. In our proposed \emph{Inter-Batch Workload Interleaving} method, each input batch $B_k$ is divided into a preparation region $P_k$ (executed on the CPU) and a kernel region $K_k$ (executed on the GPU). While $K_k$ runs GPU kernels for $B_k$, the host simultaneously performs $P_{k+1}$ for the next batch. This design decouples CPU data preparation from GPU preprocessing, allowing them to proceed in parallel. Consequently, the preparation latency of the next batch is effectively hidden behind ongoing GPU execution, thereby amortizing preprocessing overhead and enabling more flexible scheduling.
 
\subsection{Resource-aware micro-batch Schedule}
Building upon our adaptive multi-channel horizontal fusion strategy and our \emph{interleaving} strategy introduced in previous sections, we further propose a scheduling method to balance utilization at the micro-batch level. While interleaving hides preparation latency and decouples batch boundaries, the proposed scheduling method smooths per-stage load across streams to reduce pipeline bubbles during preprocessing, tiling, and decoding.  
In some scenarios, tiling and decoding involve strong data dependencies but small kernel footprints.

\begin{algorithm}[t] \small
\renewcommand{\baselinestretch}{0.95}\selectfont
\SetAlgoNlRelativeSize{-1}
\SetAlCapSkip{0.5ex}
\caption{Resource-aware Micro-batch Schedule.}
\label{alg:ra_schedule}

\KwIn{Image set $I$; warm-up stats $\Pi$; pipeline mode; baseline batch $b_0$; stream budget $P$; balance slack $\lambda$; memory cap $M_{\mathrm{cap}}$; minimum micro-batch $b_{\min}$}
\KwOut{Schedule $S=\langle S_1,\dots,S_P\rangle$; micro-batch size $m_{\text{unit}}$}
\BlankLine

\nonl\textbf{\textcolor{blue}{/* Step 1: Build candidate tasks */}}\\
$K \leftarrow \emptyset$\;
\ForEach{image $i$ in $I$}{
    $s \leftarrow \textsc{SelectTileSize}(i)$\;
    $(\mathrm{lat}, \mathrm{mem}) \leftarrow$ \\
    \quad $\textsc{PredictFromWarmup}(s,\Pi,b_0,\textsc{Mode})$\;
    push Task($i$, $s$, $\mathrm{lat}, \mathrm{mem}$) into $K$\;
}

\nonl\textbf{\textcolor{blue}{/* Step 2: Initialize streams */}}\\
\For{$p \leftarrow 1$ \KwTo $P$}{
    $S_p \leftarrow [\,]$; \quad $\mathrm{load}(p) \leftarrow 0$\;
}

\nonl\textbf{\textcolor{blue}{/* Step 3: LPT scheduling with balance check */}}\\
\While{$K$ not empty}{
    $\kappa \leftarrow$ task in $K$ with max latency\;
    Remove $\kappa$ from $K$\;
    $p^\star \leftarrow$ stream with minimum load\;
    
    \eIf{$\mathrm{load}(p^\star) + \kappa.\mathrm{lat} \le (1+\lambda)\cdot \min_q \mathrm{load}(q)$ \\ 
         \quad \textbf{and} $\textsc{MemOK}(S\cup\{\kappa\}, M_{\mathrm{cap}})$}{
        Assign $\kappa$ to stream $p^\star$\;
        $\mathrm{Load}(p^\star) \mathrel{+}= \kappa.\mathrm{lat}$\;
    }{
        $(k_1, k_2) \leftarrow \textsc{Shard}(\kappa, b_{\min})$\;
        Assign $k_1$ to stream $p^\star$\;
        $\mathrm{Load}(p^\star) \mathrel{+}= k_1.\mathrm{lat}$\;
        Push $k_2$ back into $K$\;
    }
}

\nonl\textbf{\textcolor{blue}{/* Step 4: Assign uniform micro-batch size */}}\\
$u \leftarrow$ total number of tasks across streams\;
$m_{\text{unit}} \leftarrow \max(b_{\min}, \lfloor B/u \rfloor)$\;
\ForEach{task $\kappa$ in each stream}{
    $\kappa.\mathrm{mb} \leftarrow m_{\text{unit}}$\;
}

\Return{$(S, m_{\text{unit}})$}\;
\end{algorithm}

In practice, we first profile stage latencies $(t_0,t_1,t_2)$ and per-sample memory usage during the warm-up phase of our multi-channel horizontal fusion strategy proposed in \sectionref{sec:so:sso:plp}. Then, for a micro-batch of size $m$ under a baseline batch $b_0$ and a stream allocation $(s_0,s_1,s_2)$, we formulate the predicted bottleneck cost as
\(
c(m)=
\max\Bigl\{\tfrac{t_0}{s_0},\,\tfrac{t_1}{s_1},\,\tfrac{t_2}{s_2}\Bigr\}\cdot \tfrac{m}{b_0}.
\)
Scheduling proceeds with a longest-processing-time (LPT) rule: each micro-batch is placed on the stream with the lowest accumulated load, subject to load balance and memory constraints. Low input variance further ensures schedule stability under this rule. If either constraint fails, the micro-batch is split down to a minimum size $b_{\min}$ and rescheduled. In practice, preprocessing is scheduled first, and its outputs feed into downstream queues, so that tiling and decoding automatically inherit the smoothed arrivals established by preprocessing. The global batch $B$ is evenly partitioned into micro-batches (clipped by $b_{\min}$), which amortizes the overhead of frequent launches.

As detailed in Algorithm~\ref{alg:ra_schedule}, the scheduler first \emph{builds candidate tasks}: it initializes an empty pool $K$, then for each image selects a tile size, predicts its latency and memory from warm-up statistics, and pushes a task record into $K$ (Line 2--5). It then \emph{initializes streams} by creating $P$ empty stream queues $S_p$ and zeroing their accumulated loads (Line 6--7). Next, it performs \emph{LPT scheduling with balance check}: it repeatedly pops the currently most time-consuming task $\kappa$, chooses the stream $p^\star$ with the minimum load (Line 9--13), and tests two constraints—load balance with slack $\lambda$ and global memory feasibility (Line 14). If both pass, $\kappa$ is assigned to $p^\star$ and the stream load is updated by its latency (Lines 15--16). Otherwise, $\kappa$ is \emph{sharded} at the fine granularity of the minimum micro-batch $b_{\min}$ into $(k_1,k_2)$; $k_1$ is placed onto $p^\star$ and $k_2$ is returned to the pool for later placement (Line 18--21). The loop continues until all tasks are placed. Finally, the scheduler \emph{assigns a uniform micro-batch size}: it counts the total number of scheduled tasks across streams, derives $m_{\text{unit}}=\max(b_{\min},\lfloor B/u \rfloor)$ (Line 22--23), and sets each task’s micro-batch to $m_{\text{unit}}$ (Line 24). It then returns the final stream schedule $S$ and the common micro-batch size $m_{\text{unit}}$ (Line~26).
\section{Implementation}
QRMark is built upon the Stable Signature framework while introducing an abstract and modular architecture that supports both fine-tuning and detection workflows. It adopts a multi-process scheduling design in which detector workers are launched in separate processes, enabling concurrent kernel execution on the GPU. When multiple workers share a single GPU, NVIDIA’s Multi-Process Service (MPS) can be optionally enabled to eliminate excessive context switching and improve intra-GPU parallelism. As is shown in Listing~\ref{code: QRMark example}, in practice, users can first define a custom watermarking and tiling policy to fit the texture and size of the images, then load the pretrained model and configure key runtime settings such as tiling granularity, the number of GPUs, degree of parallelism, and whether to enable Reed–Solomon decoding. QRMark supports two operating modes. The fine-tuning mode adapts the model to a target dataset and produces images embedded with watermarks. The detection mode verifies and decodes embedded watermarks from input images with efficient tile-based processing.

\begin{figure}[t]
\centering
\begin{minipage}[t]{0.45\textwidth}
\centering
\begin{lstlisting}[caption={Custom Watermarking Policy with QRMark.},label={code: QRMark example}]
import QRMark
# import other packages... e.g., PyTorch, Triton ...
# 1. Define Custom Policy Class
class CustomPolicy(QRMark.Module):
    def __init__(self, arch="sd-v2.1", bits=48):
        super().__init__()
        self.backbone = QRMark.Backbone(arch)
        self.extractor = QRMark.Extractor(bits)
    # User-defined strategy example
    def select_patch(self, x):
        p = x.unfold(2, 64, 32).unfold(3, 64, 32)
        idx = p.var((-1,-2)).sum(1).argmax(1)
        return self.gather(x, idx)
    # Define the detection forward pass
    def forward_detect(self, x):
        x_norm = QRMark.transforms.vqgan_normalize(x)
        patch = self.select_patch(x_norm)
        return self.extractor(patch, rs=True)
# 2. System Configuration
config = {"num_gpus": 4, "parallel_mode": "data_parallel", "rs": True, ...}
model = CustomPolicy().to(config)
# 3. finetune and detect
model.finetune(datasets, key=secret_key,param**)
recovered_key = model.detect(suspect_images, param**) 
accuracy = QRMark.verify(recovered_key, secret_key)
\end{lstlisting} 
\end{minipage}
\end{figure}

\section{Evaluation} \label{sec:eval}
In this section, we evaluate our QRMark system alongside several popular baseline methods and demonstrate the experimental findings. Our evaluation will focus on several aspects: (1) end-to-end system performance comparison among benchmarks, (2) adaptability to the existing watermark detection framework, (3) detection efficiency analysis, and (4) ablation study on the effects of different system design choices. This structure highlights the performance, adaptability, scalability, and expressiveness of QRMark.

{\bf Setup:}
QRMark is evaluated on a Linux server with an AMD EPYC 9534 64-Core Processor and one NVIDIA H100 GPU with 80GB HBM3 memory. The CUDA version is 12.2, and the operating system is Ubuntu 22.04. 

{\bf Baselines and Metrics:} 
Our benchmarks include two recent and representative works in diffusion model watermarking: Stable Signature~\cite{fernandez2023stable} and AquaLoRA~\cite{feng2024aqualora}. Most of QRMark’s optimizations are designed and tailored for Stable Signature. For AquaLoRA, which by default only supports a batch size of one, we minimally patched it to allow variable batch sizes during watermark extraction. To provide a comprehensive evaluation of QRMark’s performance, our metrics include: (i) detection latency and throughput, (ii) detection accuracy without watermark attacks, and (iii) robustness under various adversarial attacks. Additionally, to assess the impact of watermarking on image quality, we report image distortion using the average Peak Signal-to-Noise Ratio (PSNR).

{
{\bf Datasets:} 
During the watermark detection phase, each benchmark is evaluated on the same dataset used in its original paper. We evaluate Stable Signature using generations from prompts in the MS-COCO validation set~\cite{lin2015microsoftcococommonobjects}, following its official evaluation protocol. Using only the MS-COCO validation split ensures a fair comparison, as Stable Signature~\cite{fernandez2023stable} explicitly reports results on this set. Moreover, evaluating text-to-image models on the MS-COCO validation set is a standard practice in the literature: OpenAI generates samples using captions from the MS-COCO validation set~\cite{ramesh2022hierarchical}, and Google Brain likewise uses the COCO validation split as the standard benchmark for text-to-image evaluation~\cite{saharia2022photorealistic}. We therefore follow this established protocol. AquaLoRA is evaluated using generations from selected prompts in partiprompts~\cite{yu2022scalingautoregressivemodelscontentrich}, consistent with its original setup.
}

\subsection{End-to-End Performance}
In this section, we show the end-to-end accuracy, robustness, speedup, and latency of our QRMark on all benchmarks with various settings on a single GPU. It is important to note that QRMark uses Stable Signature as the primary baseline. The additional benchmark serves to validate the effectiveness of our algorithmic and system optimizations. 

{\bf Experiments on Stable Signature:}
Note that the distortion averages reported in the table correspond to the mean bit accuracy under the set of attacks evaluated in the respective benchmark papers. For Stable Signature, the considered attacks include \texttt{crop}, \texttt{resize}, \texttt{jpeg}, \texttt{brightness}, \texttt{contrast}, \texttt{saturation}, \texttt{sharpness}, and \texttt{overlay\_text} (full descriptions of these attacks are provided in \tabref{tab:qrmark_operation}). We adopt this collection because it represents the core set of robustness perturbations used in Stable Signature and related watermarking papers, and therefore provides a representative and comparable evaluation protocol. Importantly, the type of attack does not affect the overall efficiency of the detection pipeline, as all images pass through the same tiling, decoding, and post-processing steps regardless of the applied distortion.
Except for the case of tile size $16$, where QRMark exhibits relatively lower bit accuracy, QRMark achieves nearly identical accuracy to the original Stable Signature framework at tile sizes $32$, $48$, $64$, and $80$, and steadily approaches its robustness performance. We use a tile size of $16$ for demonstration and ablation, but such small tiles are uncommon in real deployments; larger tiles (e.g., $64$) provide substantially higher embedding capacity for the 48-bit message and therefore yield near-lossless robustness and better runtime efficiency.

\begin{figure*}[htb]
\begin{center}
\centerline{\includegraphics[width=0.99\textwidth]{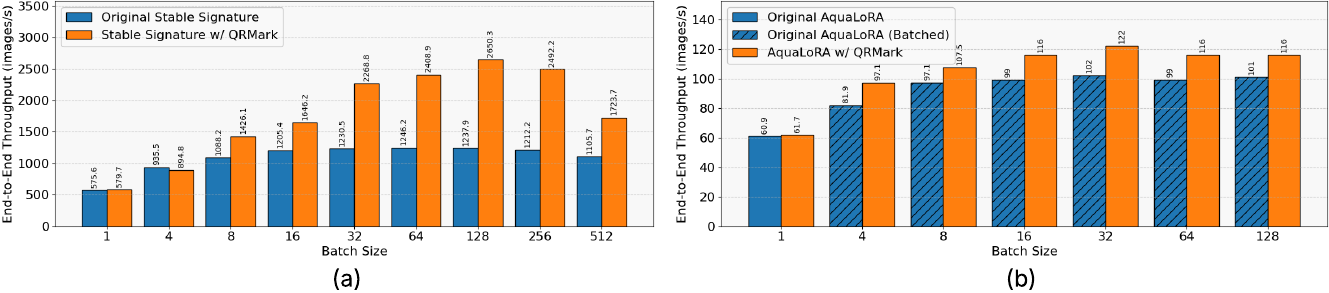}}
\vspace{-10pt}

\caption{End-to-end image throughput comparison for (a) \textit{Stable Signature} and (b) \textit{AquaLoRA} with and without our \textit{QRMark} under varying image batch sizes on a NVIDIA H100 GPU.}
\vspace{-10pt}
\label{fig:end2end_throughput}
\end{center}
\end{figure*}

\begin{figure*}[htb]
\begin{center}
\centerline{\includegraphics[width=0.99\textwidth]{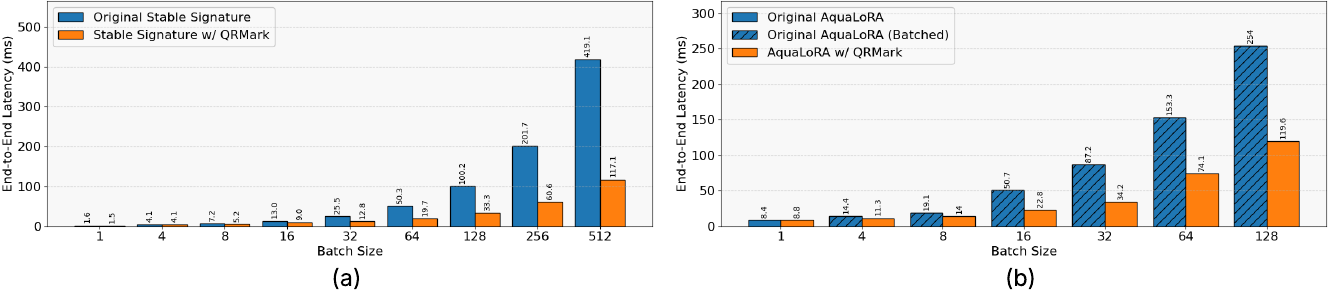}}
\vspace{-10pt}
\caption{End-to-end watermark extraction latency comparison for (a) \textit{Stable Signature} and (b) \textit{AquaLoRA} with and without our \textit{QRMark} under varying image batch sizes on a NVIDIA H100 GPU.}
\vspace{-25pt}
\label{fig:end2end_latency}
\end{center}
\end{figure*}

In \figureref{fig:end2end_throughput}, on average, QRMark achieves a $2.43\times$ throughput speedup over the original Stable Signature pipeline for image watermark detection. As for the end-to-end latency presented in \figureref{fig:end2end_latency}, as batch size increases, the batch latency of both QRMark and Stable Signature grows, but the growth rate of QRMark is significantly slower. At a batch size of 512, the latency of Stable Signature approaches nearly four times that of QRMark, highlighting the superior scalability of QRMark. The improvements in speed and latency are largely attributable to three key algorithmic and system-level optimizations: image tiling, adaptive multi-channel horizontal fusion, and resource-aware scheduling. Image tiling reduces both computational and memory costs of decoding, as each tile in our setting is only $1/16$ the size of the original image. To further optimize execution, we employ adaptive multi-channel horizontal fusion with resource-aware scheduling, which distributes CUDA streams across pipeline stages in proportion to their measured demands and assigns minibatches for each stream. This ensures that the most time-consuming stages receive more parallel capacity. The resulting multi-stream execution improves GPU utilization, boosts throughput, and lowers latency. Besides, the workload interleaving strategy, kernel fusion (full details in Appendix \ref{sec:so:kf}), and optimizations on RS Correction also contribute to the improvement.

\begin{table}[t] \small
  \centering
  \caption{Comparison between our QRMark method and baseline models on 48-bit watermarks across various tile sizes (``TS''). Adv. (Adversarial) refers to the average bit accuracy when images are under different distortions. We control the FPR at $10^{-6}$ and report the TPR. The subscript ``QR'' refers to models using our QRMark strategy using \texttt{random grid} for tiling strategy, and ``BL'' refers to baseline.}
  \vspace{3pt}
  \label{tab:overall_accuracy}
  \scalebox{0.82}{
  \begin{tabular}{clccccc}
    \toprule
    Model & TS & BitAcc. $\uparrow$ & BitAcc. (Adv.) $\uparrow$ & PSNR $\uparrow$ & TPR $\uparrow$\\
    \midrule
    \multirow{5}{*}{Stable\textsubscript{QR}}
      & 16 & 0.748 & 0.665 & 27.67 &  0.761 \\
      & 32 & 0.989 & 0.907 & 29.47 &  0.993 \\
      & 48 & 0.997 & 0.933 & 29.63 &  0.996 \\
      & 64 & 0.999 & 0.945 & 30.35 &  0.998 \\
      & 80 & 0.999 & 0.949 & 30.76 &  0.999 \\
    Stable\textsubscript{BL} & -- & 0.999 & 0.974 & 30.05 &  0.993 \\
    \midrule
      AquaLoRA\textsubscript{QR} & 256 & 0.947 & 0.883 & 17.13 &  0.970\\
      AquaLoRA\textsubscript{BL}& -- & 0.958 & 0.912 & 17.65 &  0.985\\
    \bottomrule
  \end{tabular}}
\end{table}

{\bf Experiments on AquaLoRA:}
Slightly different from how we implemented our QRMark in Stable Signature, we extended AquaLoRA by introducing the tile-based decoder only during the pre-training phase while keeping the fine-tuning process unchanged. In practice, the modified pre-training workflow mirrors that of the original AquaLoRA with three core modifications: (1) watermark-specific modifications in the encoder's forward step; (2) tile-specific adjustments allowing training of the tile-based watermark encoder and decoder; (3) hyperparameter tuning of the loss function.

On undistorted images, AquaLoRA with QRMark achieved a validation accuracy up to $94.7\%$, which is around $1\%$ difference compared to the original benchmark. To evaluate robustness, we have also included the average accuracy of watermark extraction in distorted conditions used in the original AquaLoRA paper: \texttt{colorJitter}, \texttt{crop}, \texttt{resize}, \texttt{blur}, \texttt{gaussian noise}, \texttt{jpeg}, \texttt{denoising}, and \texttt{denoising-v2}. After applying distortions to the images, QRMark achieved a less than $3\%$ performance gap to the original AquaLoRA. It suggests that QRMark has minimal impact on AquaLoRA's validation accuracy of watermark extraction and comparable robustness to the original baseline under distortions.

Moreover, as shown in \figureref{fig:end2end_throughput} and ~\ref{fig:end2end_latency}, QRMark further optimized AquaLoRA's throughput and latency by fusing the image preprocessing stages into a single CUDA kernel and reducing the intermediate global-memory writes and reads between kernels. QRMark with a batch size of 32 achieves a $2.0 \times$ speedup over the original AquaLoRA baseline and a $1.2 \times$ speedup when using the same batch size. Notably, when scaling the batch size above 128, we observed that the original AquaLoRA baseline encounters CUDA out-of-memory errors, whereas QRMark continues to sustain high throughput and reduce latency. This further highlights our method's advantage in large-batch image watermark decoding tasks.

\begin{figure}[t] \small
  \centering
  \includegraphics[width=0.9\linewidth]{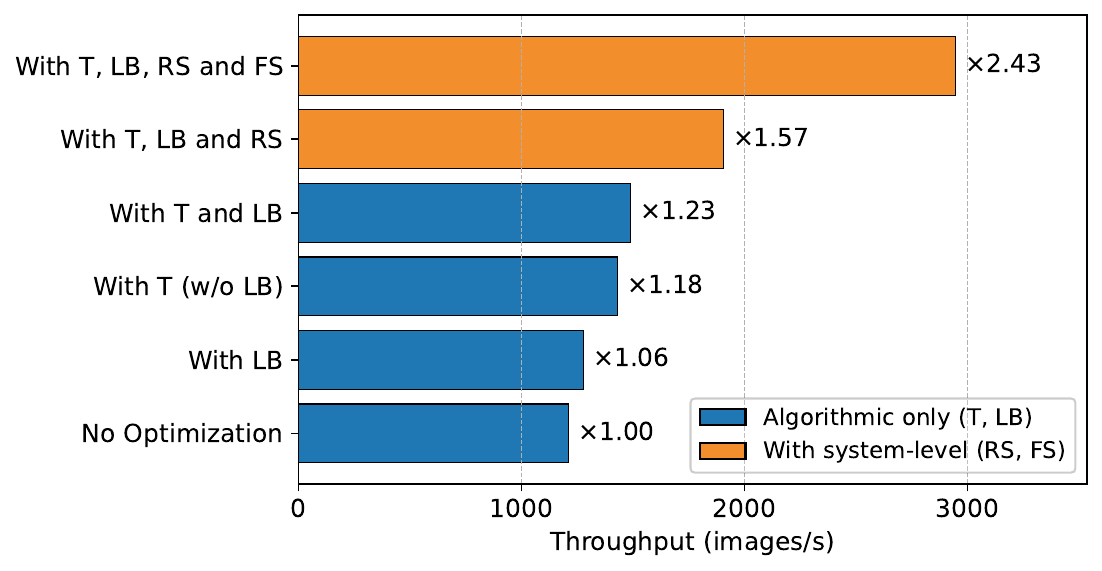}
  \vspace{-5pt}
  \caption{Throughput breakdown and incremental improvement from algorithmic and system-level optimizations. LB: large-batch processing; T: tiling; RS: optimized Reed-Solomon decoding; FS: horizontal fusion and scheduling.}
  \label{fig:breakdown}
\end{figure}

{
\subsection{Optimization Analysis} \label{sec:eval:oa}
{\bf Throughput Improvement Breakdown:}
To better illustrate the benefits of individual optimizations and quantify their contributions to end-to-end performance, we present a detailed throughput breakdown of QRMark in \figureref{fig:breakdown}. All experiments use a fixed message length of 48 bits and, when applicable, a tile size of $64$. Large-batch processing (LB) alone provides a modest $1.06\times$ improvement, while tiling (T) without LB yields a clearer benefit of $1.18\times$. Combining T and LB further increases the gain to $1.23\times$. Beyond these algorithmic optimizations, the system-level techniques deliver the dominant share of improvements: incorporating optimized Reed--Solomon decoding (RS) raises the speedup to $1.57\times$, and the addition of horizontal fusion and scheduling (FS) ultimately achieves a $2.43\times$ overall throughput improvement. These results show that although tiling and batching offer meaningful acceleration, the system-level RS and AS optimizations are the key factors driving high performance.
}

{\bf Overhead  Analysis:}
Our profiling results show that end-to-end runtime is overwhelmingly dominated by the decoding stage, which accounts for 79.0\% of total execution time. This confirms that the convolutional kernels in the extractor form the primary computational bottleneck. In comparison, preprocessing and tiling contribute 14.4\% and 6.2\%, respectively, indicating that image normalization and tile partitioning incur relatively small overheads. Synchronization and data-transfer costs constitute only 0.4\%, demonstrating that cross-stage coordination is not a limiting factor for pipeline throughput. This low synchronization cost aligns with our system design: during Reed--Solomon correction, each decoding worker transfers only a compact 48-bit message, keeping communication overhead minimal. Overall, the breakdown indicates that efficiency is constrained mainly by compute-bound decoding rather than synchronization, motivating our focus on GPU-centric and pipeline-level optimizations.

\begin{figure}[t] \small
  \centering
 \includegraphics[width=0.8\linewidth]{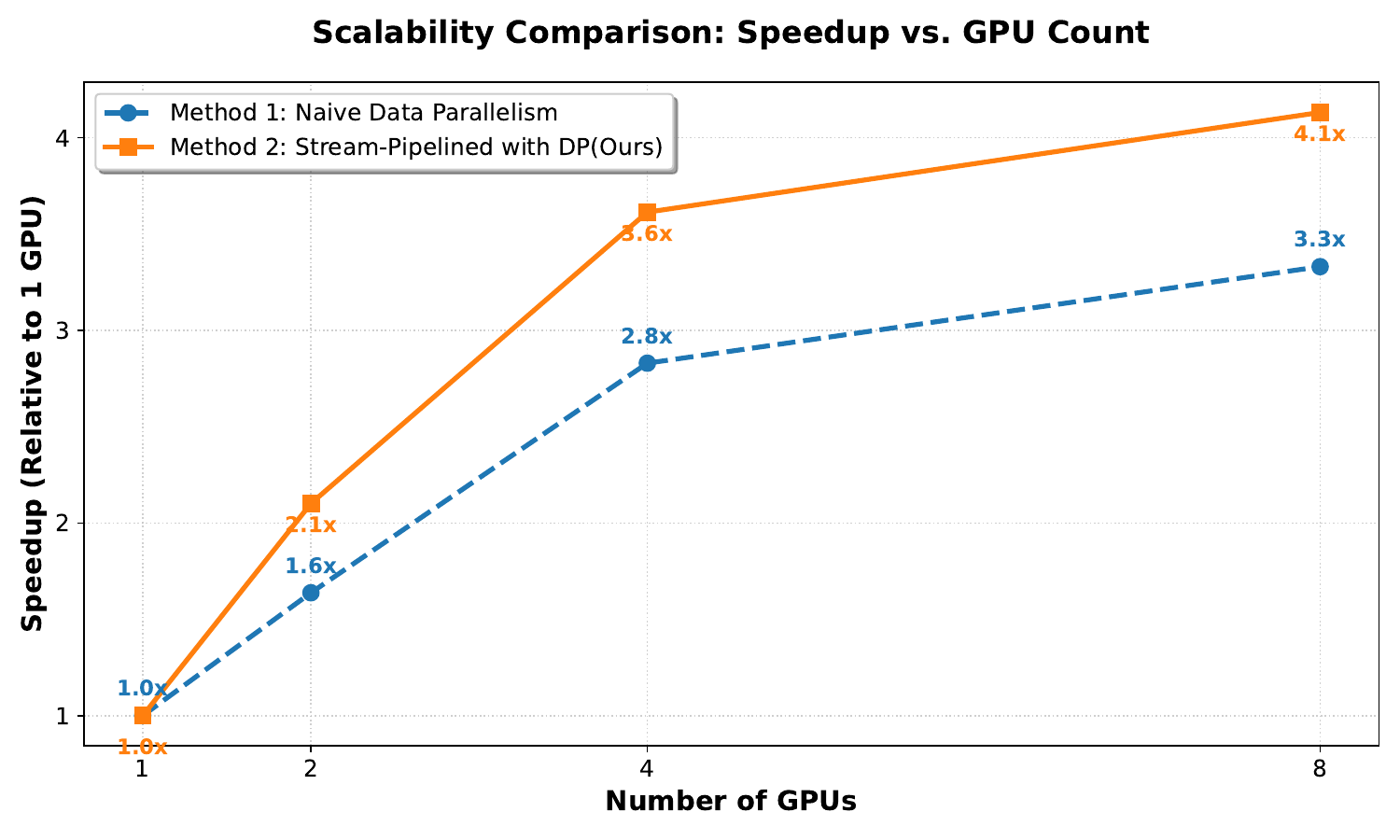}
  \vspace{-5pt}
  \caption{Scalability comparison of QRMark under data parallelism and data parallelism with horizontal fusion enabled.}
  \vspace{-5pt}
  \label{fig:scalability}
\end{figure}

{\bf Scalability Analysis:}
To highlight the scalability of QRMark under multi-GPU settings, we report throughput across different parallel configurations using a tile size of 64. As shown in \figureref{fig:scalability}, scaling with standard data parallelism exhibits inherently weak scaling for this workload. Unlike traditional DL training, the per-request computation in watermark detection is dominated by a fixed extractor that runs independently on each GPU, leaving little opportunity for cross-device parallel overlap. As a result, the baseline speedup plateaus at only $1.6\times$, $2.8\times$, and $3.3\times$ with 2, 4, and 8 GPUs, respectively. QRMark’s data-parallel design with multi-channel horizontal fusion restructures the execution pipeline to expose additional concurrency. By overlapping decoding, tiling, and Reed–Solomon correction across dedicated CUDA streams and process-level workers, the system reduces idle periods on each device and increases effective parallelism. Consequently, QRMark achieves stronger scaling, reaching $2.1\times$, $3.6\times$, and $4.1\times$ speedup at the same GPU counts—up to $30\%$ higher than the standard DP baseline. These results demonstrate that coordinated stream-level scheduling, rather than replicating the full pipeline across GPUs, is essential for converting additional hardware into meaningful throughput gains.

\section{Related Work}

\textbf{Image Watermarking:}
Image watermarking is a fundamental technique for protecting and authenticating intellectual property in image synthesis and distribution. Existing approaches can be broadly categorized into three classes: \emph{post-generation}, \emph{pre-generation}, and \emph{in-generation} watermarking~\cite{hu2024stable}. Post-generation methods~\cite{rahman2013dwt, tancik2020stegastamp, zhu2018hidden, luo2020distortion, al2007combined, jing2021hinet, zhang2020udh, zhang2019robust} embed watermarks after image creation, for instance using DWT–DCT variants, GAN-based hiding, or encoder–decoder schemes. \emph{Pre-generation} watermarking embeds watermarks into all training images~\cite{yu2021artificial, zhao2023recipe}, but this strategy incurs high computational cost when applied to large-scale diffusion models. \emph{In-generation} watermarking~\cite{fernandez2023stable, zhao2023recipe, xiong2023flexible, feng2024aqualora, wen2023tree} integrates watermark embedding into the generation process itself—for example, Stable Signature modifies the VAE~\cite{kingma2013auto} stage, while AquaLorRA and Tree-Ring embed messages directly during diffusion sampling. Owing to its strong robustness and relatively high detection efficiency, we focus on \emph{in-generation} watermarking. QRMark builds on Stable Signature and AquaLoRA, using their original implementations as baselines to demonstrate our efficiency improvements.

\textbf{Reed–Solomon Error Correction:}
Reed–Solomon (RS) codes, introduced by Reed and Solomon in 1960~\cite{reed1960polynomial}, are linear block codes over finite fields with the maximum-distance-separable (MDS) property. They are widely deployed in both consumer and mission-critical technologies, including barcodes, QR Codes, telecommunications, and broadcasting, where they ensure reliable decoding even in the presence of noise or partial corruption. In watermarking applications, RS codes strengthen robustness by encoding watermark payloads with redundancy, enabling recovery even when some image tiles are degraded by cropping, color modifications, or noise~\cite{abdul2013error,liu2025robust}. In QRMark, RS codes are employed to mitigate accuracy loss introduced by tiling.
\section{Conclusion}
This paper presents QRMark, an adaptive framework for efficient and robust embedded image watermark detection. By combining tiling-based parallel decoding, adaptive CUDA stream allocation, and Reed–Solomon error correction, QRMark significantly reduces detection latency while maintaining high watermark recoverability. Comprehensive evaluations across diverse settings demonstrate its effectiveness and robustness, establishing QRMark as a secure and scalable solution for large-scale watermark detection.

{\footnotesize \bibliographystyle{acm}
\bibliography{refs}}

@article{ramesh2022hierarchical,
  title={Hierarchical text-conditional image generation with clip latents},
  author={Ramesh, Aditya and Dhariwal, Prafulla and Nichol, Alex and Chu, Casey and Chen, Mark},
  journal={arXiv preprint arXiv:2204.06125},
  volume={1},
  number={2},
  pages={3},
  year={2022}
}

@inproceedings{rombach2022high,
  title={High-resolution image synthesis with latent diffusion models},
  author={Rombach, Robin and Blattmann, Andreas and Lorenz, Dominik and Esser, Patrick and Ommer, Bj{\"o}rn},
  booktitle={Proceedings of the IEEE/CVF conference on computer vision and pattern recognition},
  pages={10684--10695},
  year={2022}
}

@article{dhariwal2021diffusion,
  title={Diffusion models beat gans on image synthesis},
  author={Dhariwal, Prafulla and Nichol, Alexander},
  journal={Advances in neural information processing systems},
  volume={34},
  pages={8780--8794},
  year={2021}
}

@article{ho2020denoising,
  title={Denoising diffusion probabilistic models},
  author={Ho, Jonathan and Jain, Ajay and Abbeel, Pieter},
  journal={Advances in neural information processing systems},
  volume={33},
  pages={6840--6851},
  year={2020}
}

@article{song2020denoising,
  title={Denoising diffusion implicit models},
  author={Song, Jiaming and Meng, Chenlin and Ermon, Stefano},
  journal={arXiv preprint arXiv:2010.02502},
  year={2020}
}

@inproceedings{nichol2021improved,
  title={Improved denoising diffusion probabilistic models},
  author={Nichol, Alexander Quinn and Dhariwal, Prafulla},
  booktitle={International conference on machine learning},
  pages={8162--8171},
  year={2021},
  organization={PMLR}
}

@article{saharia2022photorealistic,
  title={Photorealistic text-to-image diffusion models with deep language understanding},
  author={Saharia, Chitwan and Chan, William and Saxena, Saurabh and Li, Lala and Whang, Jay and Denton, Emily L and Ghasemipour, Kamyar and Gontijo Lopes, Raphael and Karagol Ayan, Burcu and Salimans, Tim and others},
  journal={Advances in neural information processing systems},
  volume={35},
  pages={36479--36494},
  year={2022}
}

@article{rahman2013dwt,
  title={A DWT, DCT and SVD based watermarking technique to protect the image piracy},
  author={Rahman, Md Maklachur},
  journal={arXiv preprint arXiv:1307.3294},
  year={2013}
}

@article{zhang2019robust,
  title={Robust invisible video watermarking with attention},
  author={Zhang, Kevin Alex and Xu, Lei and Cuesta-Infante, Alfredo and Veeramachaneni, Kalyan},
  journal={arXiv preprint arXiv:1909.01285},
  year={2019}
}

@inproceedings{zhu2018hidden,
  title={Hidden: Hiding data with deep networks},
  author={Zhu, Jiren and Kaplan, Russell and Johnson, Justin and Fei-Fei, Li},
  booktitle={Proceedings of the European conference on computer vision (ECCV)},
  pages={657--672},
  year={2018}
}

@misc{kingma2013auto,
  title={Auto-encoding variational bayes},
  author={Kingma, Diederik P and Welling, Max and others},
  year={2013},
  publisher={Banff, Canada}
}

@article{feng2024aqualora,
  title={Aqualora: Toward white-box protection for customized stable diffusion models via watermark lora},
  author={Feng, Weitao and Zhou, Wenbo and He, Jiyan and Zhang, Jie and Wei, Tianyi and Li, Guanlin and Zhang, Tianwei and Zhang, Weiming and Yu, Nenghai},
  journal={arXiv preprint arXiv:2405.11135},
  year={2024}
}

@article{wen2023tree,
  title={Tree-ring watermarks: Fingerprints for diffusion images that are invisible and robust},
  author={Wen, Yuxin and Kirchenbauer, John and Geiping, Jonas and Goldstein, Tom},
  journal={arXiv preprint arXiv:2305.20030},
  year={2023}
}

@inproceedings{fernandez2023stable,
  title={The stable signature: Rooting watermarks in latent diffusion models},
  author={Fernandez, Pierre and Couairon, Guillaume and J{\'e}gou, Herv{\'e} and Douze, Matthijs and Furon, Teddy},
  booktitle={Proceedings of the IEEE/CVF International Conference on Computer Vision},
  pages={22466--22477},
  year={2023}
}

@article{zhang2020udh,
  title={Udh: Universal deep hiding for steganography, watermarking, and light field messaging},
  author={Zhang, Chaoning and Benz, Philipp and Karjauv, Adil and Sun, Geng and Kweon, In So},
  journal={Advances in Neural Information Processing Systems},
  volume={33},
  pages={10223--10234},
  year={2020}
}

@inproceedings{luo2020distortion,
  title={Distortion agnostic deep watermarking},
  author={Luo, Xiyang and Zhan, Ruohan and Chang, Huiwen and Yang, Feng and Milanfar, Peyman},
  booktitle={Proceedings of the IEEE/CVF conference on computer vision and pattern recognition},
  pages={13548--13557},
  year={2020}
}

@inproceedings{yu2021artificial,
  title={Artificial fingerprinting for generative models: Rooting deepfake attribution in training data},
  author={Yu, Ning and Skripniuk, Vladislav and Abdelnabi, Sahar and Fritz, Mario},
  booktitle={Proceedings of the IEEE/CVF International conference on computer vision},
  pages={14448--14457},
  year={2021}
}

@inproceedings{tancik2020stegastamp,
  title={Stegastamp: Invisible hyperlinks in physical photographs},
  author={Tancik, Matthew and Mildenhall, Ben and Ng, Ren},
  booktitle={Proceedings of the IEEE/CVF conference on computer vision and pattern recognition},
  pages={2117--2126},
  year={2020}
}

@article{al2007combined,
  title={Combined DWT-DCT digital image watermarking},
  author={Al-Haj, Ali},
  journal={Journal of computer science},
  volume={3},
  number={9},
  pages={740--746},
  year={2007}
}

@inproceedings{jing2021hinet,
  title={Hinet: Deep image hiding by invertible network},
  author={Jing, Junpeng and Deng, Xin and Xu, Mai and Wang, Jianyi and Guan, Zhenyu},
  booktitle={Proceedings of the IEEE/CVF international conference on computer vision},
  pages={4733--4742},
  year={2021}
}

@article{hu2024stable,
  title={Stable signature is unstable: removing image watermark from diffusion models},
  author={Hu, Yuepeng and Jiang, Zhengyuan and Guo, Moyang and Gong, Neil},
  journal={arXiv preprint arXiv:2405.07145},
  year={2024}
}

@inproceedings{xiong2023flexible,
  title={Flexible and secure watermarking for latent diffusion model},
  author={Xiong, Cheng and Qin, Chuan and Feng, Guorui and Zhang, Xinpeng},
  booktitle={Proceedings of the 31st ACM International Conference on Multimedia},
  pages={1668--1676},
  year={2023}
}

@article{zhao2023recipe,
  title={A recipe for watermarking diffusion models},
  author={Zhao, Yunqing and Pang, Tianyu and Du, Chao and Yang, Xiao and Cheung, Ngai-Man and Lin, Min},
  journal={arXiv preprint arXiv:2303.10137},
  year={2023}
}

@article{reed1960polynomial,
  title={Polynomial codes over certain finite fields},
  author={Reed, Irving S and Solomon, Gustave},
  journal={Journal of the society for industrial and applied mathematics},
  volume={8},
  number={2},
  pages={300--304},
  year={1960},
  publisher={SIAM}
}

@misc{lin2015microsoftcococommonobjects,
      title={Microsoft COCO: Common Objects in Context}, 
      author={Tsung-Yi Lin and Michael Maire and Serge Belongie and Lubomir Bourdev and Ross Girshick and James Hays and Pietro Perona and Deva Ramanan and C. Lawrence Zitnick and Piotr Dollár},
      year={2015},
      eprint={1405.0312},
      archivePrefix={arXiv},
      primaryClass={cs.CV},
      url={https://arxiv.org/abs/1405.0312}, 
}

@article{redmark,
  author       = {Mahdi Ahmadi and
                  Alireza Norouzi and
                  S. M. Reza Soroushmehr and
                  Nader Karimi and
                  Kayvan Najarian and
                  Shadrokh Samavi and
                  Ali Emami},
  title        = {ReDMark: Framework for Residual Diffusion Watermarking on Deep Networks},
  journal      = {CoRR},
  volume       = {abs/1810.07248},
  year         = {2018},
  url          = {http://arxiv.org/abs/1810.07248},
  eprinttype    = {arXiv},
  eprint       = {1810.07248},
  bibsource    = {dblp computer science bibliography, https://dblp.org}
}

@inproceedings{narayanan2019pipedream,
  title={PipeDream: Generalized pipeline parallelism for DNN training},
  author={Narayanan, Deepak and Harlap, Aaron and Phanishayee, Amar and Seshadri, Vivek and Devanur, Nikhil R and Ganger, Gregory R and Gibbons, Phillip B and Zaharia, Matei},
  booktitle={Proceedings of the 27th ACM symposium on operating systems principles},
  pages={1--15},
  year={2019}
}

@inproceedings{tan2019efficientnet,
  title={Efficientnet: Rethinking model scaling for convolutional neural networks},
  author={Tan, Mingxing and Le, Quoc},
  booktitle={International conference on machine learning},
  pages={6105--6114},
  year={2019},
  organization={PMLR}
}

@inproceedings{chen2016xgboost,
  title={Xgboost: A scalable tree boosting system},
  author={Chen, Tianqi and Guestrin, Carlos},
  booktitle={Proceedings of the 22nd acm sigkdd international conference on knowledge discovery and data mining},
  pages={785--794},
  year={2016}
}

@inproceedings{wang2025sleepermark,
  title={Sleepermark: Towards robust watermark against fine-tuning text-to-image diffusion models},
  author={Wang, Zilan and Guo, Junfeng and Zhu, Jiacheng and Li, Yiming and Huang, Heng and Chen, Muhao and Tu, Zhengzhong},
  booktitle={Proceedings of the Computer Vision and Pattern Recognition Conference},
  pages={8213--8224},
  year={2025}
}

@inproceedings{wang2024rap,
  title={Rap: Resource-aware automated gpu sharing for multi-gpu recommendation model training and input preprocessing},
  author={Wang, Zheng and Wang, Yuke and Deng, Jiaqi and Zheng, Da and Li, Ang and Ding, Yufei},
  booktitle={Proceedings of the 29th ACM International Conference on Architectural Support for Programming Languages and Operating Systems, Volume 2},
  pages={964--979},
  year={2024}
}

@misc{czolbe2021lossfunctiongenerativeneural,
      title={A Loss Function for Generative Neural Networks Based on Watson's Perceptual Model}, 
      author={Steffen Czolbe and Oswin Krause and Ingemar Cox and Christian Igel},
      year={2021},
      eprint={2006.15057},
      archivePrefix={arXiv},
      primaryClass={cs.LG},
      url={https://arxiv.org/abs/2006.15057}, 
}

@misc{loshchilov2019decoupledweightdecayregularization,
      title={Decoupled Weight Decay Regularization}, 
      author={Ilya Loshchilov and Frank Hutter},
      year={2019},
      eprint={1711.05101},
      archivePrefix={arXiv},
      primaryClass={cs.LG},
      url={https://arxiv.org/abs/1711.05101}, 
}

@misc{YouTubeHowItWorksSearch,
  title  = {YouTube Search — How YouTube Works},
  author = {{YouTube}},
  year   = {2025},
  note   = {“With over 500 hours of content uploaded to YouTube every minute …”},
  howpublished = {\url{https://www.youtube.com/intl/en_us/howyoutubeworks/product-features/search/}}
}

@article{Steel2025TikTokHour,
  title   = {Just Another Hour on TikTok: Reverse-engineering unique identifiers to obtain a complete slice of TikTok},
  author  = {Steel, Benjamin and Schirmer, Miriam and Ruths, Derek and Pfeffer, Juergen},
  journal = {arXiv preprint arXiv:2504.13279},
  year    = {2025},
  note    = {Estimates \~117M posts in a single day},
  url     = {https://arxiv.org/abs/2504.13279}
}

@article{Wang2015FaceSearchAtScale,
  title   = {Face Search at Scale: 80 Million Gallery},
  author  = {Wang, Dayong and Otto, Charles and Jain, Anil K.},
  journal = {arXiv preprint arXiv:1507.07242},
  year    = {2015},
  note    = {Cites Facebook white paper: 250B photos total; 350M new photos/day (2013)},
  url     = {https://arxiv.org/abs/1507.07242}
}

@article{abdul2013error,
  title={Error correcting codes for robust color wavelet watermarking},
  author={Abdul, Wadood and Carr{\'e}, Philippe and Gaborit, Philippe},
  journal={EURASIP Journal on Information Security},
  volume={2013},
  number={1},
  pages={1},
  year={2013},
  publisher={Springer}
}

@article{liu2025robust,
  title={Robust and secure image watermarking algorithm resisting JPEG compression and cropping based on Reed Solomon (RS) codes},
  author={Liu, Yushi and Shi, Canghong and Hu, Xuefei and Qi, Chao and Niu, Xianhua},
  journal={Signal, Image and Video Processing},
  volume={19},
  number={3},
  pages={253},
  year={2025},
  publisher={Springer}
}

@techreport{lim1978decoding,
  title={A decoding procedure for the Reed-Solomon codes},
  author={Lim, Raymond S},
  year={1978}
}

@inproceedings{min2024watermark,
  title={A watermark-conditioned diffusion model for ip protection},
  author={Min, Rui and Li, Sen and Chen, Hongyang and Cheng, Minhao},
  booktitle={European Conference on Computer Vision},
  pages={104--120},
  year={2024},
  organization={Springer}
}

@inproceedings{xu2009tiling,
  title={Tiling for performance tuning on different models of GPUs},
  author={Xu, Chang and Kirk, Steven R and Jenkins, Samantha},
  booktitle={2009 Second International Symposium on Information Science and Engineering},
  pages={500--504},
  year={2009},
  organization={IEEE}
}

@inproceedings{rolih2024divide,
  title={Divide and conquer: High-resolution industrial anomaly detection via memory efficient tiled ensemble},
  author={Rolih, Bla{\v{z}} and Ameln, Dick and Vaidya, Ashwin and Akcay, Samet},
  booktitle={Proceedings of the IEEE/CVF Conference on Computer Vision and Pattern Recognition},
  pages={3866--3875},
  year={2024}
}

@inproceedings{jangda2020model,
  title={Model-based warp overlapped tiling for image processing programs on GPUs},
  author={Jangda, Abhinav and Guha, Arjun},
  booktitle={Proceedings of the ACM International Conference on Parallel Architectures and Compilation Techniques},
  pages={317--328},
  year={2020}
}

@inproceedings{ozge2019power,
  title={The power of tiling for small object detection},
  author={Ozge Unel, F and Ozkalayci, Burak O and Cigla, Cevahir},
  booktitle={Proceedings of the IEEE/CVF conference on computer vision and pattern recognition workshops},
  pages={0--0},
  year={2019}
}

@misc{yu2022scalingautoregressivemodelscontentrich,
      title={Scaling Autoregressive Models for Content-Rich Text-to-Image Generation}, 
      author={Jiahui Yu and Yuanzhong Xu and Jing Yu Koh and Thang Luong and Gunjan Baid and Zirui Wang and Vijay Vasudevan and Alexander Ku and Yinfei Yang and Burcu Karagol Ayan and Ben Hutchinson and Wei Han and Zarana Parekh and Xin Li and Han Zhang and Jason Baldridge and Yonghui Wu},
      year={2022},
      eprint={2206.10789},
      archivePrefix={arXiv},
      primaryClass={cs.CV},
      url={https://arxiv.org/abs/2206.10789}, 
}

@article{begum2020digital,
  title={Digital image watermarking techniques: a review},
  author={Begum, Mahbuba and Uddin, Mohammad Shorif},
  journal={Information},
  volume={11},
  number={2},
  pages={110},
  year={2020},
  publisher={MDPI}
}

@misc{tensorrt,
  title        = {NVIDIA TensorRT: Programmable Inference Accelerator},
  author       = {{NVIDIA Corporation}},
  howpublished = {\url{https://developer.nvidia.com/tensorrt}},
  year         = {2023},
  note         = {Accessed: 2025-08-20}
}

@inproceedings{li2024distrifusion,
  title={Distrifusion: Distributed parallel inference for high-resolution diffusion models},
  author={Li, Muyang and Cai, Tianle and Cao, Jiaxin and Zhang, Qinsheng and Cai, Han and Bai, Junjie and Jia, Yangqing and Li, Kai and Han, Song},
  booktitle={Proceedings of the IEEE/CVF Conference on Computer Vision and Pattern Recognition},
  pages={7183--7193},
  year={2024}
}

@article{kodaira2023streamdiffusion,
  title={Streamdiffusion: A pipeline-level solution for real-time interactive generation},
  author={Kodaira, Akio and Xu, Chenfeng and Hazama, Toshiki and Yoshimoto, Takanori and Ohno, Kohei and Mitsuhori, Shogo and Sugano, Soichi and Cho, Hanying and Liu, Zhijian and Keutzer, Kurt},
  journal={arXiv preprint arXiv:2312.12491},
  year={2023}
}

@article{wadhera2022comprehensive,
  title={A comprehensive review on digital image watermarking},
  author={Wadhera, Shweta and Kamra, Deepa and Rajpal, Ankit and Jain, Aruna and Jain, Vishal},
  journal={arXiv preprint arXiv:2207.06909},
  year={2022}
}

@article{zhong2023brief,
  title={A brief, in-depth survey of deep learning-based image watermarking},
  author={Zhong, Xin and Das, Arjon and Alrasheedi, Fahad and Tanvir, Abdullah},
  journal={Applied Sciences},
  volume={13},
  number={21},
  pages={11852},
  year={2023},
  publisher={MDPI}
}

@inproceedings{hussain2022faststamp,
  title={Faststamp: Accelerating neural steganography and digital watermarking of images on fpgas},
  author={Hussain, Shehzeen and Sheybani, Nojan and Neekhara, Paarth and Zhang, Xinqiao and Duarte, Javier and Koushanfar, Farinaz},
  booktitle={Proceedings of the 41st IEEE/ACM International Conference on Computer-Aided Design},
  pages={1--9},
  year={2022}
}

@article{zhang2025robust,
  title={Robust and secure code watermarking for large language models via ml/crypto codesign},
  author={Zhang, Ruisi and Javidnia, Neusha and Sheybani, Nojan and Koushanfar, Farinaz},
  journal={arXiv preprint arXiv:2502.02068},
  year={2025}
}

@inproceedings{uchida2017embedding,
  title={Embedding watermarks into deep neural networks},
  author={Uchida, Yusuke and Nagai, Yuki and Sakazawa, Shigeyuki and Satoh, Shin'ichi},
  booktitle={Proceedings of the 2017 ACM on international conference on multimedia retrieval},
  pages={269--277},
  year={2017}
}

@inproceedings{wang2025timestep,
  title={Timestep-Aware Diffusion Model for Extreme Image Rescaling},
  author={Wang, Ce and Hu, Zhenyu and Sun, Wanjie and Chen, Zhenzhong},
  booktitle={Proceedings of the IEEE/CVF International Conference on Computer Vision},
  pages={15594--15603},
  year={2025}
}

@article{ding2024patched,
  title={Patched denoising diffusion models for high-resolution image synthesis},
  author={Ding, Zheng and Zhang, Mengqi and Wu, Jiajun Wu and Tu, Zhuowen},
  year={2024},
  publisher={ICLR}
}

@inproceedings{zhang2023sine,
  title={Sine: Single image editing with text-to-image diffusion models},
  author={Zhang, Zhixing and Han, Ligong and Ghosh, Arnab and Metaxas, Dimitris N and Ren, Jian},
  booktitle={Proceedings of the IEEE/CVF conference on computer vision and pattern recognition},
  pages={6027--6037},
  year={2023}
}

@inproceedings{madar2025tiled,
  title={Tiled Diffusion},
  author={Madar, Or and Fried, Ohad},
  booktitle={Proceedings of the Computer Vision and Pattern Recognition Conference},
  pages={7795--7804},
  year={2025}
}

@article{sun2025patchedserve,
  title={PATCHEDSERVE: A Patch Management Framework for SLO-Optimized Hybrid Resolution Diffusion Serving},
  author={Sun, Desen and Zhao, Zepeng and Wang, Yuke},
  journal={arXiv preprint arXiv:2501.09253},
  year={2025}
}

\clearpage
	\appendix
\section{Optimization and design in QRMark}

\subsection{Kernel Fusion for Preprocessing}\label{sec:so:kf}
To fully exploit GPU resources, we focus on the CPU-bound preprocessing stage as a key overhead. The goal is to consolidate the entire image transformation pipeline into a single GPU kernel, thereby reducing redundant memory movement and kernel launches.
Concretely, we fuse the sequence of operations (\textit{Raw Image $\rightarrow$ Resize $\rightarrow$ CenterCrop $\rightarrow$ Normalize}) into one Triton kernel. The fused operator directly consumes raw image buffers and integrates resizing, cropping, type conversion, and per-channel scale–bias normalization within a unified affine mapping. Unlike the earlier design with separate tensor operations, this kernel executes the full transformation in a single pass. It is compiled with Triton’s autotuning facilities, which explore block size and warp count to select an efficient configuration. Once tuned, the configuration is cached and reused across subsequent calls. In practice, this fused kernel acts as a drop-in replacement, providing lower latency and higher throughput. We further evaluate its benefits in \sectionref{sec:eval:oa}.

\subsection{ML-based Tile Size Predictor}\label{subsec:ml_tile_predictor}
In \sectionref{sec:alo}, we illustrate the pipeline for image watermark detection. However, in real‑world scenarios, a significant challenge arises: given a target image, how can we determine its watermark’s tile size? A naive strategy might involve sliding tiles of varying sizes over the image and processing each through multiple decoders pretrained for specific tile sizes. Such an approach, however, incurs prohibitive computational overhead and is impractical for real‑time deployment. To address this challenge, we propose a ML‑based Watermark Tile‑Size Predictor that efficiently estimates the watermark tile size from a given image in a single forward pass. Our predictor combines convolutional neural network feature extraction with gradient‑boosted regression, selecting an EfficientNet \cite{tan2019efficientnet} + XGBoost \cite{chen2016xgboost} architecture for its balance of performance and interoperability. Concretely, we adopt a pretrained EfficientNet as a backbone to extract a compact feature vector from each input image. These feature embeddings are then fed into an XGBoost regressor, which predicts the tile size with high accuracy. Similar to our preprocessing latency predictor, both training‑data collection and model training are conducted offline, thereby eliminating runtime profiling overhead.

\subsection{Ablation Study on QRMark Robustness}\label{sec:eval:ab}

\begin{table}[t]
  \centering
  \caption{\raggedright Validation accuracy w/ different tiling strategies.}
  \vspace{2px}
  \label{tab:sampling_vs_tile}
  \scalebox{0.8}{ 
  \begin{tabular}{lccccc}
    \toprule
    Tiling Strategy & 16 & 32 & 48 & 64 & 80 \\
    \midrule
    Random      & 0.93 & 0.99 & 0.99 & 0.99 & 0.99 \\
    Random Grid & 0.75 & 0.97 & 0.99 & 0.99 & 0.99 \\
    Fixed       & 0.81 & 0.84 & 0.85 & 0.99 & 0.99 \\
    \bottomrule
  \end{tabular}
  }
\end{table}

\begin{table}[t] \small
  \centering
  \caption{\raggedright Validation accuracy under different watermark \#bits.}
  \label{tab:acc_vs_bits}
  \scalebox{0.95}{
  \begin{tabular}{lccccccc}
    \toprule
    \#bits & 40 & 48 & 56 & 64 & 72 & 80 & 96 \\
    \midrule
    Bit Acc.  & 0.99 & 0.99 & 0.98 & 0.91 & 0.89 & 0.84 & 0.77 \\
    Word Acc. & 0.99 & 0.99 & 0.29 & 0.00 & 0.00 & 0.00 & 0.00 \\
    \bottomrule
  \end{tabular}
  }
  \vspace{-10pt}
\end{table}

\begin{table}[t] \small
  \centering
  \caption{Bit accuracy under various tiling strategies and attacks. ``C-0.1'': \texttt{crop 0.1}, ``C-0.5'': \texttt{crop 0.5}, ``R-0.5'': \texttt{resize 0.5}, ``BL'': \texttt{blur}, ``BR-2'': \texttt{brightness 2}, ``CON-2'': \texttt{contrast 2}.}
  \vspace{2px}
  \label{tab:tile_attack_accuracy}
  \scalebox{0.82}{
  \begin{tabular}{lccccccc}
    \toprule
    Tiling Strategy & None & C-0.1 & C-0.5 & R-0.5 & BL & BR-2 & CON-2 \\
    \midrule
    Random      & 0.99 & 0.99 & 0.99 & 0.88 & 0.79 & 0.92 & 0.95 \\
    Random Grid & 0.99 & 0.99 & 0.94 & 0.94 & 0.85 & 0.99 & 0.98 \\
    Fixed       & 0.99 & 0.99 & 0.99 & 0.67 & 0.80 & 0.87 & 0.94 \\
    \bottomrule
  \end{tabular}
  }
\end{table}

We conduct ablation studies to better understand the design choices in QRMark. These experiments aim to identify suitable trade-offs that maximize detection efficiency and message capacity while keeping accuracy degradation within acceptable limits. The design choices examined include tile size, message length, and tiling strategies.

\noindent\textbf{Tiling Strategies under Adversarial Attacks:}
We evaluate three tiling strategies in \tabref{tab:qrmark_operation} for sampling tiles from images. As shown in \tabref{tab:tile_attack_accuracy}, all three strategies achieve near-perfect accuracy in the absence of attacks or cropping. Under stronger perturbations, however, \texttt{Random Grid} exhibits superior robustness against resizing, brightness, and contrast distortions, while \texttt{Fixed} suffers substantial degradation under resizing. In contrast, \texttt{Random} is more resilient to resizing but less stable under blur. These findings motivate our adoption of \texttt{Random Grid} as the default tiling strategy for QRMark, offering both robustness and efficiency.

\noindent\textbf{Tiling Strategies with Different Tile Sizes:}
To further assess the impact of different tiling strategies, we measure detection accuracy for 48-bit watermarks across varying tile sizes (\tabref{tab:sampling_vs_tile}). Both \texttt{Random} and \texttt{Random Grid} achieve consistently high accuracy once the tile size reaches $32 \times 32$. In contrast, \texttt{Fixed} performs poorly at smaller tile sizes and only approaches comparable accuracy when the tile size is increased to 64 or larger. Taken together, the results in \tabref{tab:tile_attack_accuracy} and \tabref{tab:sampling_vs_tile} clearly highlight the superior performance of \texttt{Random Grid}, establishing it as the most robust and reliable tiling strategy for QRMark across a broad range of settings.

\noindent\textbf{Detection Accuracy Across Watermark Bit Lengths:}
To evaluate the adaptability of QRMark to different payload sizes, we measure detection accuracy for watermark lengths ranging from 40 to 96 bits (\tabref{tab:acc_vs_bits}). Bit-level accuracy remains nearly perfect up to 64 bits but then declines sharply, dropping to 0.89 at 72 bits and below 0.80 at 96 bits. In contrast, word-level accuracy degrades much earlier, collapsing beyond 48 bits—falling to 0.29 at 56 bits and reaching 0 at 64 bits and above. This divergence clearly indicates that, while individual bits can still be recovered with only moderate fidelity, symbol-level decoding becomes increasingly unreliable once redundancy is insufficient to correct accumulated errors. Based on this analysis, we conservatively set 64 bits as the maximum payload size for a $64 \times 64$ tile and adopt 48 bits as the default message length, thereby striking a balance between detection accuracy and embedding capacity in QRMark.

\section{Details on Reed-Solomon correction} \label{sec:rs}
\subsection{Notation:}
Let~$GF(2^{m})$ be the Galois field with~$m$‐bit symbols.
We denote the code parameters by
$(n,k,t)$, where
$n$ is the codeword length, $k$ the number of information symbols,
and $t=\lfloor\frac{n-k}{2}\rfloor$ the maximum number of
correctable symbol errors.
A length-$n$ vector~$C=(C_0,\dots,C_{n-1})\in GF(2^{m})^n$
represents one codeword, while the binary representation
contains $n\cdot m$ bits.

\subsection{Systematic Evaluation-Based Encoding:}
In this example, let $\mathcal X=\{X_0,\dots,X_{n-1}\}\subset GF(2^{m})$
be $n$ fixed, pairwise-distinct evaluation points
(the {\em evaluation set}).
We first split the $k\cdot m$ message bits into
$k$~symbols $M=(M_0,\dots,M_{k-1})\in GF(2^{m})^k$,
and associate them with the first $k$ evaluation points:
$(X_i,M_i)$ for $i=0,\dots,k-1$.

\begin{enumerate}[label=(\alph*),nosep,leftmargin=1.2em]
  \item \textbf{Interpolation.}\;%
        By Lagrange interpolation, a unique polynomial
        $P(x)\in GF(2^{m})[x]$ of degree $\deg P<k$ satisfies
        \begin{equation}
            P(X_i)=M_i,\qquad i=0,\dots,k-1 .
        \end{equation}
        We implement interpolation explicitly via the basis polynomials
        $$
          \ell_i(x)=\Bigl(\prod_{j\neq i}\frac{x-X_j}{X_i-X_j}\Bigr),\qquad
          P(x)=\sum_{i=0}^{k-1} M_i\,\ell_i(x).
        $$
        In our implementation, the above is realized in $\mathcal O(k^2)$ operations.

  \item \textbf{Evaluation.}\;%
        Evaluating $P(x)$ on the entire set~$\mathcal X$ yields the systematic codeword
        \begin{equation}
          C_i=P(X_i),\qquad i=0,\dots,n-1 ,
        \end{equation}
        where $C_i=M_i$ for $i<k$. Conversion back to bits is symbol-wise and leaves
        the first $k\cdot m$ bits unchanged.
\end{enumerate}

\noindent
The auxiliary routine
\verb|_scale_polynomial(poly,scalar)|
performs coefficient-wise scaling in~$ GF(2^{m})$ and is
required by both steps.

\subsection{Berlekamp–Welch Decoding:} \label{sec:rs:dec}
Given a received word
$R=(R_0,\dots,R_{n-1})$ that differs from~$C$
in at most~$t$ symbols,
the Berlekamp–Welch (B–W) procedure finds the original
polynomial~$P(x)$ without recourse to the Berlekamp–Massey chain:

\begin{enumerate}[label=(\alph*),nosep,leftmargin=1.2em]
  \item \textbf{Error-locator and numerator.}\;%
        Seek two polynomials
        \begin{align}
          Q(x) &\not\equiv 0,\quad \deg Q \le t,\\[-0.25ex]
          N(x) &\phantom{\not\equiv 0{}}\quad\deg N \le t+k-1,
        \end{align}
        such that
        \begin{equation}\label{eq:bw-system}
          N(X_i)=R_i\,Q(X_i),\qquad i=0,\dots,n-1 .
        \end{equation}
        Eq.~\eqref{eq:bw-system} forms a homogeneous linear system in the coefficients
        of $Q$ and $N$ and is solved via Gaussian elimination over $GF(2^{m})$
        (time complexity $\mathcal O(n^3)$; smaller in practice).

  \item \textbf{Message polynomial.}\;%
        Provided the solution satisfies $Q(x)\not\equiv 0$ and
        \mbox{$\deg N<\deg Q + k$}, we recover
        \begin{equation}
           P(x)=N(x)\,/\,Q(x).
        \end{equation}

  \item \textbf{Symbol recovery.}\;%
        Because we employ a systematic encoder, the message is read by evaluation:
        \begin{equation}
           \hat M_i = P(X_i),\qquad i=0,\dots,k-1 .
        \end{equation}
        No coefficient extraction is necessary, which keeps the implementation confined
        to the \verb|encode|/\verb|decode| routines.
\end{enumerate}

\begin{algorithm}[t]
\DontPrintSemicolon
\SetAlgoLined
\caption{\textsc{RS\_Encode} — Systematic Reed--Solomon}
\KwIn{
  $\mathbf m$ -- bitstring of length $k\!\times\!m$ \;
  $(n,k,t)$ -- code parameters, $t=\lfloor(n-k)/2\rfloor$ \;
  $GF(2^{m})$ -- operating finite field \;
  $\mathcal X=\{X_0,\dots,X_{n-1}\}$ -- $n$ distinct evaluation points
}

\KwOut{Codeword $\mathbf C=(C_0,\dots,C_{n-1})\in GF(2^{m})^{n}$}

\textbf{1. Bit\,→\,Symbol conversion}:\;
\hspace*{1.5em}$M \leftarrow (M_0,\dots,M_{k-1})$ 

\textbf{2. Interpolate message polynomial $P(x)$}:\;
\hspace*{1.5em}\For{$i\gets0$ \KwTo $k-1$}{
  build basis $\displaystyle\ell_i(x)=\prod_{j\neq i}\frac{x-X_j}{X_i-X_j}$\;
}
\hspace*{1.5em}$P(x)\leftarrow\displaystyle\sum_{i=0}^{k-1} M_i\,\ell_i(x)$  

\textbf{3. Generate systematic codeword}:\;
\hspace*{1.5em}\For{$i\gets0$ \KwTo $n-1$}{
  $C_i \leftarrow P(X_i)$ 
}

\textbf{4. Return} $\mathbf C$
\end{algorithm}

The decoder returns
(i)~the corrected $k\cdot m$ message bits,
(ii)~the full $n\cdot m$ codeword bits, and
(iii) the number of symbol errors actually corrected,
allowing downstream components to gauge confidence.

A $(n,k)$ Reed–Solomon code over $ GF(2^{m})$
obeys the classical length bound
\begin{equation}
   n_{\max}=2^{m}-1 .
\end{equation}
Each symbol carries $m$ bits; hence the {\em maximum}
binary payload is $n_{\max}\,m$.
For our application, we adopt two concrete field sizes:

\begin{itemize}[nosep,leftmargin=1.2em]
  \item {\bf $m=4$ (GF(16)).}\;
        $n_{\max}=15$ symbols $\Rightarrow 60$ bits.
        Choosing $k=12$ provides exactly $48$ information bits
        and leaves $t=\frac{15-12}{2}=1.5\!\xrightarrow{}\!1$
        a symbol of correction, which empirical tests show
        is sufficient for our task.
  \item {\bf $m=8$ (GF(256)).}\;
        $n_{\max}=255$ symbols, matching the {\it de-facto} standard
        for larger payloads.  Here, $k$ is selected dynamically
        to balance capacity and robustness in long messages.
\end{itemize}
    
\end{document}